%% file: paper2.tex
\documentclass[twocolumn]{aastex63}
\shorttitle{Sgr B1}
\shortauthors{Simpson et al.}

\usepackage{latexsym}
\usepackage{graphicx}
\usepackage{amssymb}
\usepackage{amsmath}
\usepackage{longtable}
\usepackage{epsf}

\begin{document}

\title{
Sagittarius B1 - A Patchwork of H II Regions and PhotoDissociation Regions
}

\author[0000-0001-8095-4610]{Janet P. Simpson}
\affiliation{SETI Institute \\
189 Bernardo Ave. \\
Mountain View, CA 94043, USA}
\email{janet.p.simpson@gmail.com}

\author{Sean W. J. Colgan}
\affiliation{NASA Ames Research Center \\
MS 245-6 \\
Moffett Field, CA 94035-1000}

\author{Angela S. Cotera}
\affiliation{SETI Institute \\
189 Bernardo Ave. \\
Mountain View, CA 94043, USA}

\author[0000-0002-2521-1985]{Michael J. Kaufman}
\affiliation{San Jose State University \\
Department of Physics and Astronomy, One Washington Square \\ 
San Jose, CA 95192-0106}

\author{Susan R. Stolovy}
\affiliation{El Camino College \\
Physics Department, 16007 Crenshaw Blvd. \\ 
Torrance, CA 90506}

\begin{abstract}  

Sgr~B1 is a luminous \ion{H}{2} region in the Galactic Center immediately next to the massive star-forming giant molecular cloud Sgr~B2 and apparently connected to it from their similar radial velocities.
In 2018 we showed from SOFIA FIFI-LS observations of the [\ion{O}{3}] 52 and 88 \micron\ lines that there is no central exciting star cluster and that the ionizing stars must be widely spread throughout the region.
Here we present SOFIA FIFI-LS observations of the [\ion{O}{1}] 146 and [\ion{C}{2}] 158 \micron\ lines formed in the surrounding photodissociation regions (PDRs).
We find that these lines correlate neither with each other nor with the [\ion{O}{3}] lines although together they correlate better with the 70 \micron\ {\it Herschel} PACS images from Hi-GAL.
We infer from this that Sgr~B1 consists of a number of smaller \ion{H}{2} regions plus their associated PDRs, some seen face-on and the others seen more or less edge-on.
We used the PDR Toolbox to estimate densities and the far-ultraviolet intensities exciting the PDRs. 
Using models computed with Cloudy, we demonstrate possible appearances of edge-on PDRs and show that the density difference between the PDR densities and the electron densities estimated from the [\ion{O}{3}] line ratios is incompatible with pressure equilibrium unless there is a substantial pressure contribution from either turbulence or magnetic field or both.
We also conclude that the hot stars exciting Sgr~B1 are widely spaced throughout the region at substantial distances from the gas with no evidence of current massive star formation.

\end{abstract}



\section{Introduction}

A photodissociation region (PDR) is usually described as the interface between 
an \ion{H}{2} region surrounding a massive star and a nearby molecular cloud.
More generally, a PDR can be thought of as any diffuse interstellar gas 
that is heated by photons with energies less than 
needed to ionize hydrogen (ionization potential $IP = 13.6$ eV). 
Notable markers of PDRs in the interstellar medium (ISM) 
are strong lines of elements with $IP < 13.6$~eV, 
such as C$^+$ and Si$^+$
(Tielens \& Hollenbach 1985; Bennett et al. 1994). 
A detailed analysis of PDRs heated by luminous external sources 
was first performed by Tielens \& Hollenbach (1985) 
and expanded to lower densities and lower luminosity heating sources 
by Wolfire et al. (1990) and Kaufman et al. (1999). 
In these papers, they computed the chemistry and line emission for gases
heated by input far-ultraviolet (FUV, 6 -- 13.6~eV) photons. 
Their range of hydrogen nucleus density $n$ went 
from $10^2$ to $10^6$~cm$^{-3}$ 
and their range of FUV energy intensities $G_0$ went from 1 to $10^6$, 
where they defined $G_0$ as the ratio of the photon intensity 
divided by the local interstellar FUV intensity 
($1.3 \times 10^{-4}$ erg~cm$^{-2}$~s$^{-1}$~sr$^{-1}$, Habing 1968). 
In addition to lines of [\ion{C}{2}] and [\ion{Si}{2}] (Table~1), 
features that characterize PDRs are 
lines from neutral species like [\ion{O}{1}] and [\ion{C}{1}], 
molecular hydrogen, and CO, and emission from warm dust. 
An additional interesting PDR component 
is the multi-atom polycyclic aromatic hydrocarbon (PAH) molecule's numerous 
near and mid-infrared (NIR, MIR) bands (Tielens 2008).  

\input tab1.tex

The PDR analysis software referenced above 
is most useful for PDRs surrounding stars 
that do not produce photons energetic enough to ionize hydrogen,
such as young stellar objects (YSOs) or reflection nebulae 
(e.g., Sandell et al. 2015 and references therein; 
Bernard-Salas et al. 2015).
For hotter stars, it is informative to use a code that first produces 
an \ion{H}{2} region and then uses the remaining non-hydrogen-ionizing photons 
to compute the properties of the more distant PDR. 
These codes include the follow-on to the Tielens \& Hollenbach (1985) code 
described by Kaufman et al. (2006) 
and the gaseous nebula code Cloudy (Ferland et al. 2017; Abel et al. 2005; 
Shaw et al. 2005).
We shall employ both of these in this paper. 

The Galactic Center (GC, distance $\sim 8.2$~kpc, Abuter et al. 2019; Reid et al. 2019) is notable for its massive 
$4 \times 10^6$ M$_\odot$ black hole Sgr~A*, massive young star clusters, 
and dense molecular clouds (Morris \& Serabyn 1996).  
The GC gas is observed to be significantly warmer than the GC dust 
(e.g., Rodr\'iguez-Fern\'andez et al. 2004) --- 
this could be due to significant turbulence (e.g., Ginsburg et al. 2016) 
or to cosmic-ray heating, as seen in H$_3^+$ (Oka et al. 2019).
Substantial magnetic fields are inferred from 
the observed nonthermal radio emission (see the review of Ferri\`ere 2009)
and suggested equipartition with the cosmic rays (Oka et al. 2019). 
The GC also includes the results of massive star formation: 
the nuclear stellar cluster in Sgr A, the nearby Arches and Quintuplet clusters 
(ages 3.5 and 4.8 Myr, Schneider et al. 2014),
and the most active star-forming region of the Galaxy, Sgr~B2. 

A good overview image of the GC is given by Hankins et al. (2020), 
whose Galactic Center Legacy Survey  
mapped the warm dust seen at 25 and 37~\micron\ in the brightest parts 
($\pm \sim 0.7\degr$ from Sgr~A*)
with the Faint Object infraRed CAmera for the SOFIA Telescope 
(FORCAST; Herter et al. 2012) on the 
Stratospheric Observatory for Infrared Astronomy
(SOFIA; Young et al. 2012; Temi et al. 2014).
Other comprehensive images are given by Yusef-Zadeh et al. (2004) 
and Heywood et al. (2019), showing both the thermal and non-thermal emission at radio wavelengths, 
and the far-infrared (FIR) images from {\it Herschel Space Observatory} of Molinari et al. (2011). In the NIR, 
PAH emission is seen everywhere in the images 
that Stolovy et al. (2006) took with 
the Infrared Array Camera (IRAC, Fazio et al. 2004) on 
the {\it Spitzer Space Telescope} (Werner et al. 2004). 
The GC emits a plethora of ionized lines, mostly indicating low excitation 
\ion{H}{2} regions and associated PDRs 
(Simpson 2018). 
PDR lines observed in the GC include the [\ion{C}{2}] 158~\micron\ line
(Langer et al. 2017; Garc\'ia et al. 2016; Rodr\'iguez-Fern\'andez et al. 2004), 
[\ion{O}{1}] (Iserlohe et al. 2019; Rodr\'iguez-Fern\'andez et al. 2004), 
various lines of H$_2$ (Mills et al. 2017; Rodr\'iguez-Fern\'andez et al. 2001), 
 and [\ion{C}{1}] and CO lines (Martin et al. 2004).

The velocity structure of the GC is consistent with 
four streams of gas on open orbits around the Sgr~A nucleus (Kruijssen et al. 2015; for further discussion, see  Tress et al. 2020, who model the effects of bars and feedback on the velocity structure of the GC gas).
In their theory, 
star formation occurs when the gas is compressed near the orbit pericenter 
(Longmore et al. 2013; Kruijssen et al. 2015; Barnes et al. 2017). 
On the other hand, Sormani et al. (2020) find that star formation occurs 
in the dust lanes that result from gas flows in the Galactic bar. 
Both theories result in the stars forming 
in a ring of gas orbiting $\sim 100 -200$~pc from Sgr~A* 
(e.g., Dale et al. 2019; Sormani et al. 2020).
Sgr~B2 and the Sgr~B giant molecular cloud are located at the positive Galactic longitude end 
of the orbit with the youngest time since pericenter passage (Barnes et al. 2017), 
with the much older Arches and Quintuplet Clusters forming on an earlier orbit 
(orbital azimuthal period $\sim 3.69$~Myr, Kruijssen et al. 2015).

\begin{figure*}
\plotone{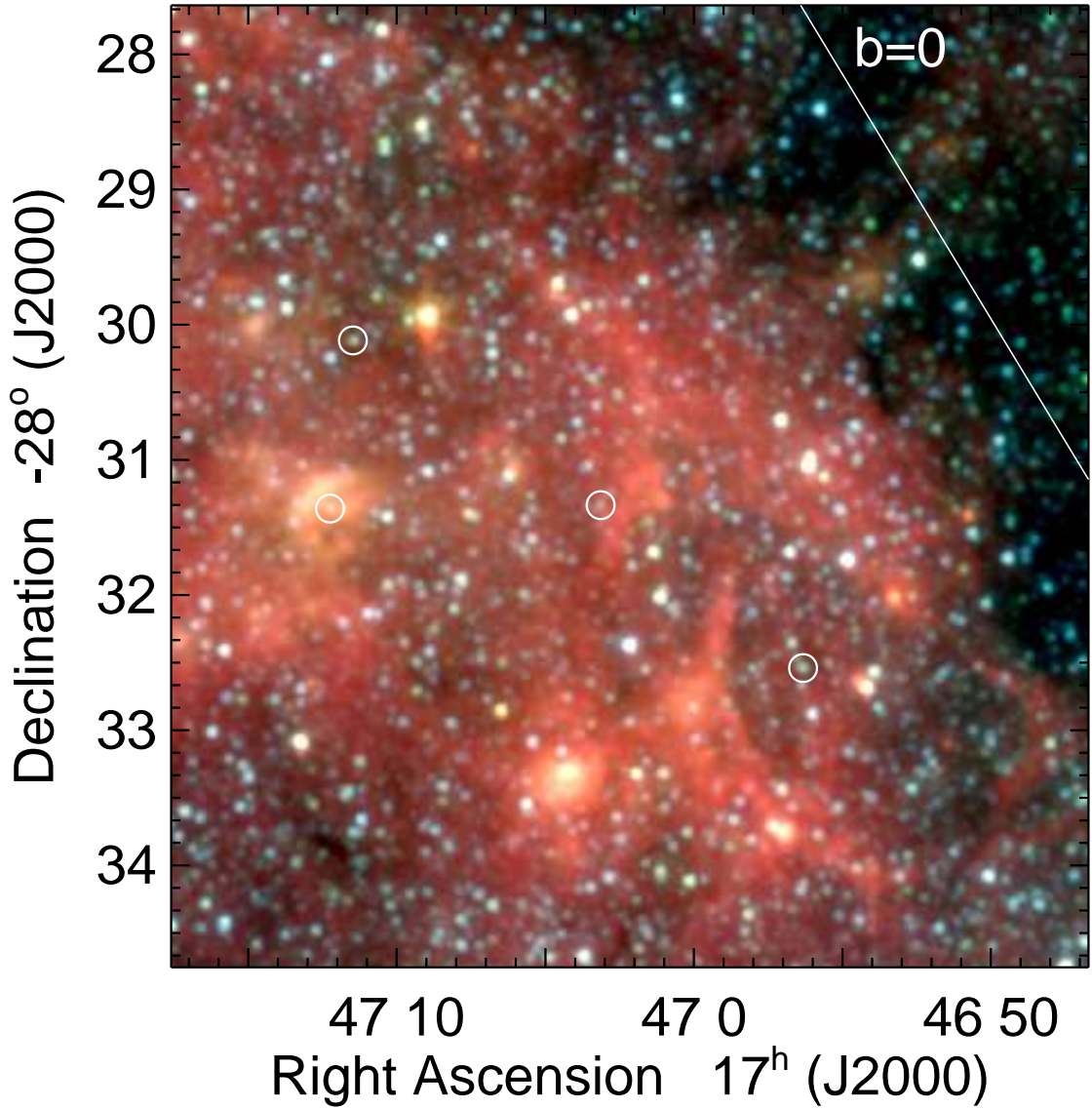}
\caption{Three-color image of Sgr~B1 from IRAC bands 1, 2, and 4 (blue: 3.6~\micron, green: 4.5~\micron, and red: 8.0~\micron) 
from Stolovy et al. (2006), plotted on a logarithmic scale.
The four encircled stars show the locations of the O supergiant and the Wolf-Rayet stars identified by Mauerhan et al. (2010).
The saturated red blob at 
R.A. (J2000) 17$^{\rm h}$47$^{\rm m}$09\fs0 decl. (J2000) $-$28\degr~30\arcmin~00\arcsec\ 
is the OH/IR star OH 0.55-0.06 (2MASS J17470898-2829561) that was studied by Oka et al. (2019).
 The location of Galactic latitude $b = 0$ is marked. 
}
\end{figure*}

Sgr~B1 (G0.50$-$0.05) is the luminous \ion{H}{2} region immediately closer to Sgr~A  
than the heavily extincted Sgr~B2 (G0.70$-$0.05).  
Its radial velocity ($V_{\rm LSR} \sim 45$~km s$^{-1}$, Downes et al. 1980) 
is consistent with it being part of the same Sgr~B molecular cloud as Sgr~B2. 
However, its actual location in the GC is unclear and difficult to determine.
Although prominent in ionized gas (e.g., Mehringer et al. 1992; Yusef-Zadeh et al. 2004) 
and warm dust (Hankins et al. 2020), Sgr~B1 is barely noticeable 
in either molecular gas (e.g., Henshaw et al. 2016) or 
very cold dust (e.g., Arendt et al. 2019; Battersby et al. 2020). 
In particular, there is no correlation of the warm dust of Sgr~B1,
seen with SOFIA FORCAST at 25 and 37~\micron\ (Hankins et al. 2020) 
or at 70~\micron\ with {\it Herschel} PACS (Molinari et al. 2011, 2016) 
with any emission from the lines that could represent the molecular-cloud part of a Sgr~B1 PDR, 
such as CO (Tanaka et al. 2009; Martin et al. 2004) 
or [\ion{C}{1}] 609~\micron\ (Martin et al. 2004). 
The stream of gas that contains Sgr~B2 lies at higher Galactic latitude 
than Sgr~B1 --- 
this stream is readily visible in the longer wavelength {\it Herschel} SPIRE images 
of Molinari et al. (2011) and the 870~\micron\ APEX images of Immer et al. (2012),
who describe it as a `Dust Ridge' lying at Galactic latitude $b \sim 0$, 
or northwest of Sgr~B1.
However, although its velocity and Galactic longitude should place Sgr~B1 
on the far side of the gas-stream orbit as viewed from the earth, 
Sgr~B1 has sufficiently low extinction compared to Sgr~B2 
that it is easily seen in the NIR. 
In addition, 
it does not suffer the extinction on its west side that one would expect 
for objects that lie behind the low-Galactic-latitude edge of the Dust Ridge,
but the east side of Sgr~B1 shows significantly more MIR extinction (Simpson 2018) 
as well as formaldehyde absorption at the velocity of Sgr~B1 (Mehringer et al. 1995).

Figure~1 is a 3-color image of Sgr~B1 taken with {\it Spitzer} IRAC 
by Stolovy et al. (2006).
Although IRAC bands 1 (3.6~\micron) and 2 (4.5~\micron) mostly show just stars, 
IRAC band 4 (8.0~\micron) is dominated by the emission from PAHs 
(Draine \& Li 2007). 
A few areas in Sgr~B1 also show extra emission in the IRAC bands   
due to warm dust that is heated by fairly high FUV intensities 
(Arendt et al. 2008). 
In general, the location of the 8~\micron\ emission agrees quite well 
with the high-spatial-resolution 8.4~GHz radio emission 
observed by Mehringer et al. (1992), 
from which we infer that the ionized gas 
and the PAH molecules are co-located.
Good agreement is also seen with the warm dust  
imaged at 70~\micron\ (Molinari et al. 2011, 2016) 
and the SOFIA FORCAST images of A. Cotera et al. (in preparation).

Although Sgr~B1 is thought to be part of the Sgr~B molecular cloud 
with its active star-forming region Sgr~B2, 
the immediate impression on looking at Figure~1  
is that there is no indication 
of any central star cluster that could ionize the gas.
Instead, there are a few bright regions that may be dense knots of gas, 
but the morphology of most of the region looks more like the rims of 
ionized bubbles of gas and dust with low-density gas between them
(e.g., Churchwell et al. 2006).
These rim structures are not correlated with any gas of higher density 
as determined from the ratio of the [\ion{O}{3}] 52/88~\micron\ lines 
observed with SOFIA by Simpson et al. (2018).
In such rim structures, 
the increased FIR and radio brightness 
is due to longer pathlengths through the gas and dust and not increased density.

Simpson et al. (2018) also found that the peaks of the relatively high excitation O$^{++}$ ions 
are widely scattered across the region and not in any central cluster.  
As a result, they suggested that the ionizing stars did not form in situ,
but drifted into the Sgr~B molecular cloud in their orbits around the GC.
If so, these stars would be several million years old, such as the 
Wolf-Rayet and O supergiants observed by Mauerhan et al. (2010) 
and plotted in Figure~1.

In this paper, we analyze the PDRs that surround the Sgr~B1 \ion{H}{2} regions
with the goal of determining their structure and relation 
to the associated \ion{H}{2} regions. 
The lines discussed in this paper are listed in Table~1.
Section~2 describes observations of Sgr~B1 taken 
in the [\ion{C}{2}] 158 and [\ion{O}{1}] 146~\micron\ lines 
with the Field Imaging Far-Infrared Line Spectrometer 
(FIFI-LS; Colditz et al. 2018; Fischer et al. 2018) on SOFIA.
Section~3 describes the results and how they compare to observations 
of the warm and cool dust and ionized gas.   
Section~4 compares the observed line intensities, including those of the [\ion{O}{3}] lines
measured by Simpson et al. (2018), to models of \ion{H}{2} regions and PDRs, 
and Section~5 presents the summary and conclusions. 

\section{Observations}

\begin{figure*}
\centering
\plottwo{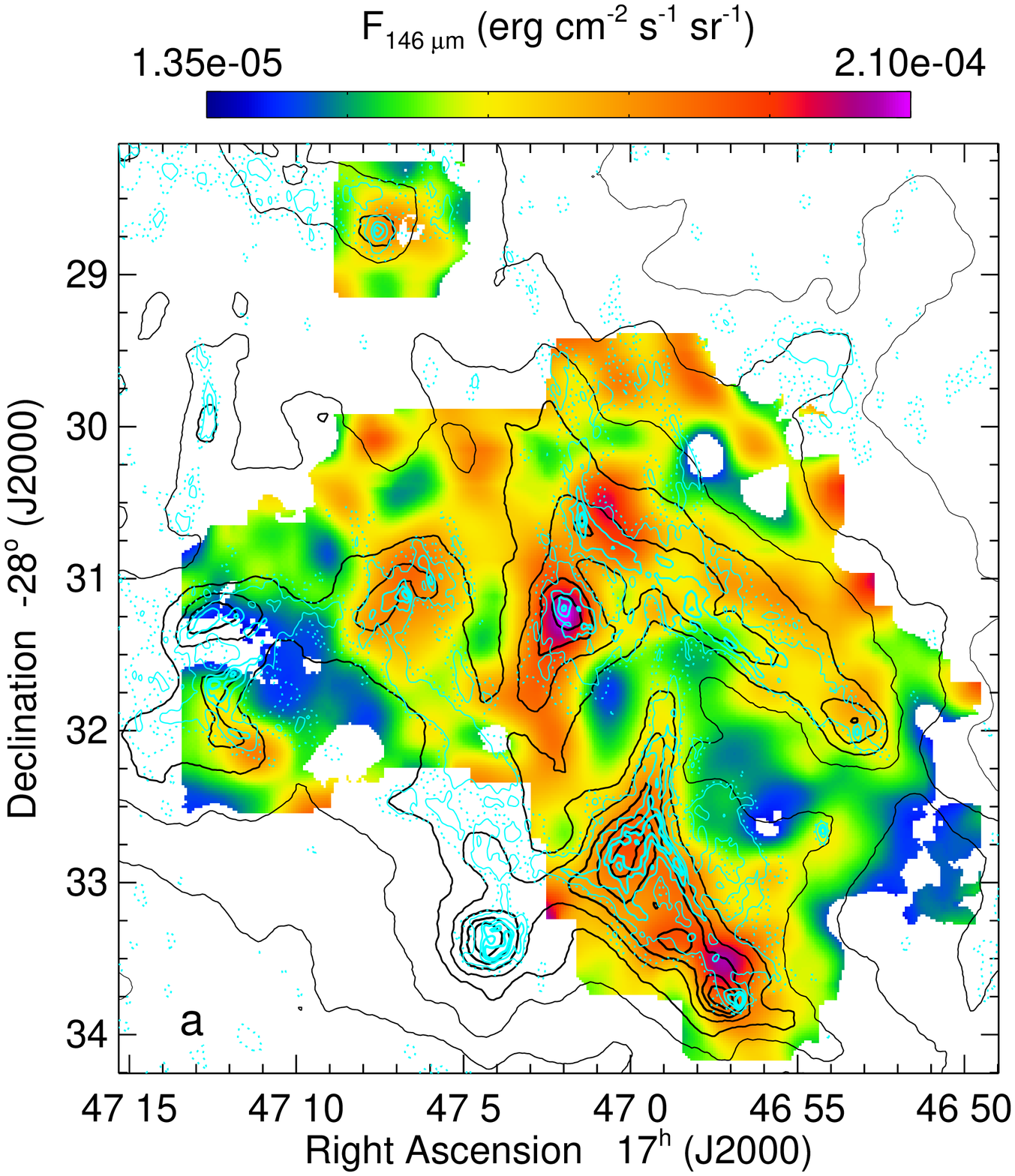}{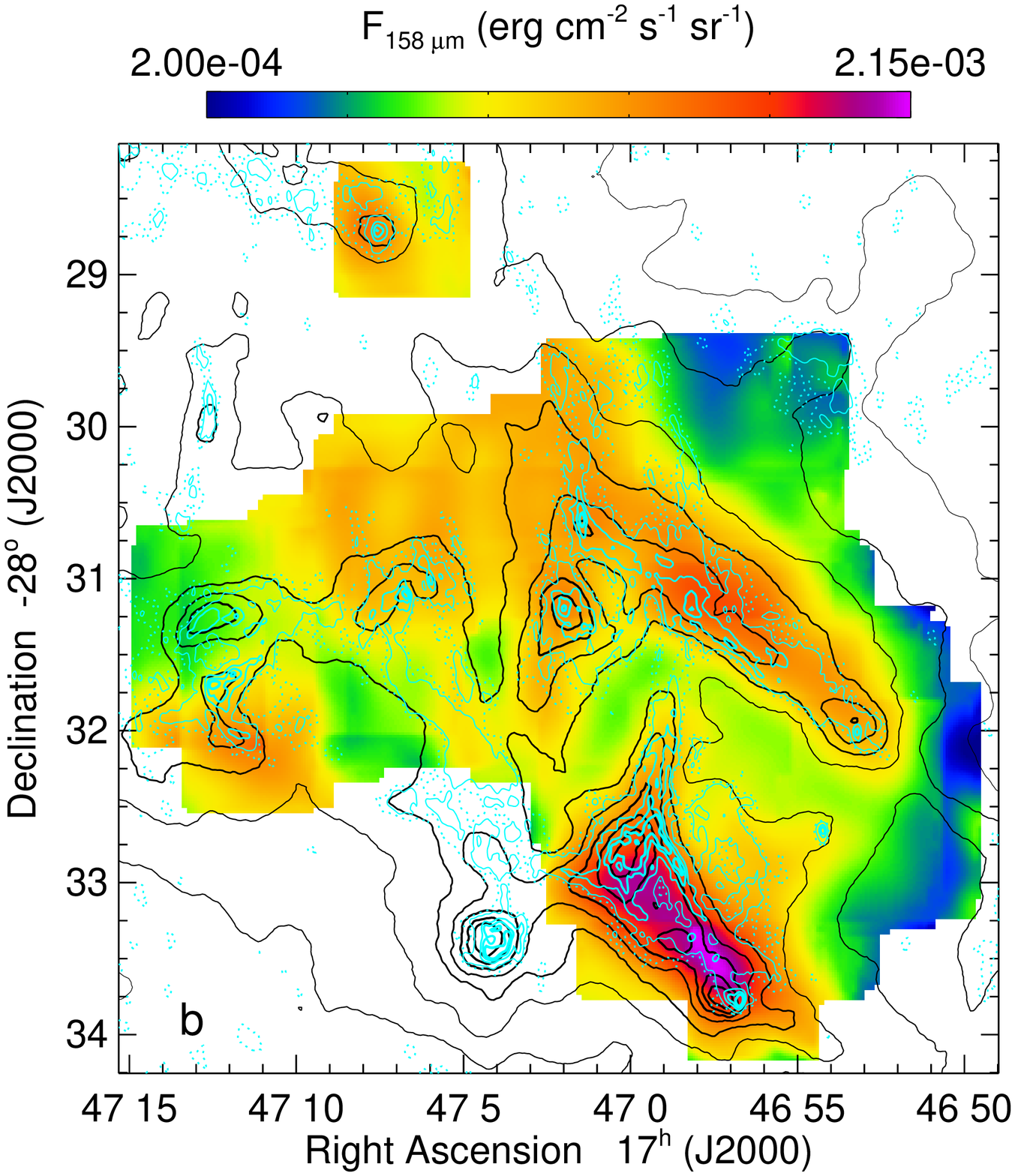}
\caption{The observed intensities of the [\ion{O}{1}] 146 and [\ion{C}{2}] 158~\micron\ lines.
The black contours are the 70~\micron\ Herschel PACS image (Molinari et al. 2016) 
and the cyan contours are the 8.4~GHz VLA image of Mehringer et al. (1992).
(a) The [\ion{O}{1}] 146~\micron\ line (maximum intensity $= 2.10 \times 10^{-4}$~erg~cm$^{-2}$~s$^{-1}$~sr$^{-1}$).
The holes are pixels with signal/noise $ < 5$.
(b) The [\ion{C}{2}] 158~\micron\ line (maximum intensity $= 2.15 \times 10^{-3}$~erg~cm$^{-2}$~s$^{-1}$~sr$^{-1}$).
}
\end{figure*}

We mapped Sgr~B1 with SOFIA FIFI-LS in 2016 and 2017, 
with SOFIA flying from Christchurch, NZ. 
FIFI-LS has two spectrometers 
(`blue channel', 51 -- 120~\micron, and `red channel', 115--200~\micron), 
which operate simultaneously. 
The details of the observations and data reduction 
for the blue-channel [\ion{O}{3}] lines at 52 and 88~\micron\ 
were described by Simpson et al. (2018). 
These details also pertain to the red-channel lines 
of [\ion{O}{1}] 146~\micron\ and [\ion{C}{2}] 158~\micron\  
that are discussed in this paper. 
The main differences with the blue-channel data reduction 
are that the red-channel pixels are 12\arcsec\ instead of 
6\arcsec\ for the blue-channel pixels and,  
as produced by the pipeline, 
the pixels in the output are 2\arcsec\ instead of 1\arcsec.
However, after the line intensities of both channels were measured, 
the maps of the blue-channel lines were padded and the red-channel maps were resampled,
so that all four line maps have 1\arcsec\ pixels and the same coordinates.
The intensities of the [\ion{O}{1}] 146 and [\ion{C}{2}] 158~\micron\ lines are plotted 
in Figure~2. 
FITS files of all four lines are included in this paper. 

The chopper throws were 4\arcmin\ in 2016 and $\sim 6\arcmin$ in 2017, 
the purpose of the small throw being to minimize coma in the blue-channel lines. 
However, there is substantial extended line and continuum emission 
in the Sgr~B region at 146--158~\micron\ that is not found at 88~\micron, 
such that some of this emission was in the reference beams 
and was subtracted in each chop/nod.
As a result, the overall levels of the line intensity are slightly different 
in the parts of each line image that were observed in 2016
compared to the parts observed in 2017 
(the continuum intensity levels are too different for the continua to be useful). 
These junctions are visible in images of the line intensities, which are plotted in Figure~2.

\section{Results}

\begin{figure*}
\includegraphics[width=140mm]{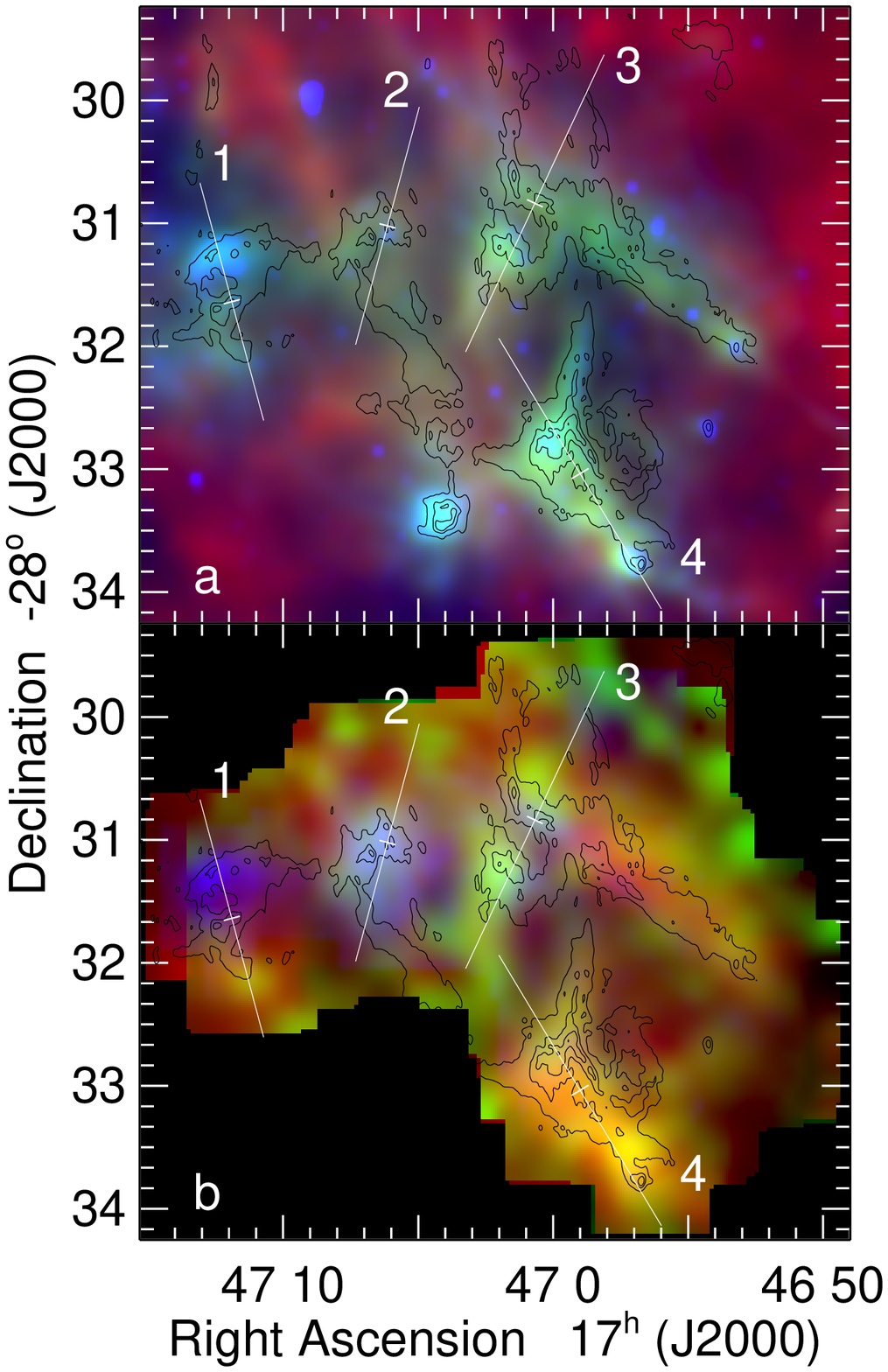}
\caption{
Three-color images of Sgr~B1. The black contours are the 8.4~GHz radio image 
of Mehringer et al. (1992).
The positions of the four artificial slits of Figure~4 are shown; 
the slit centers are marked by a short bar. 
(a) Continuum images. Blue is the IRAC 8~\micron\ image of Figure 1,
green is the {\it Herschel} PACS 70~\micron\ image, 
and red is the {\it Herschel} PACS 160~\micron\ image (Molinari et al. 2016), 
all plotted on a linear scale.
(b) Line images from our FIFI-LS data. 
Blue is the [\ion{O}{3}] 88~\micron\ image from Simpson et al. (2018),
and green and red are the [\ion{O}{1}] 146~\micron\ and [\ion{C}{2}] 158~\micron\ images, 
respectively, plotted on a linear scale. 
}
\end{figure*}

\begin{figure*}
\plottwo{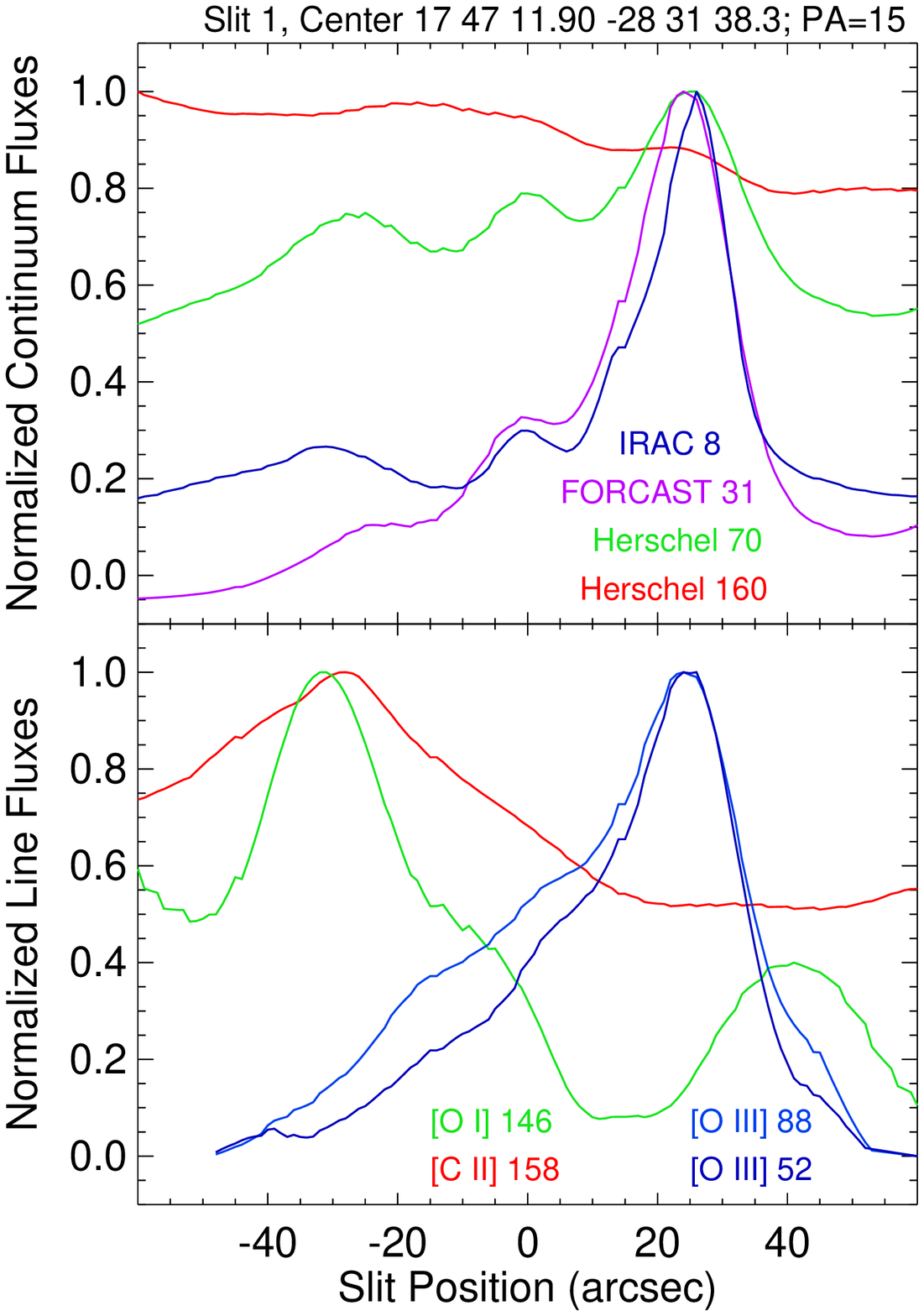}{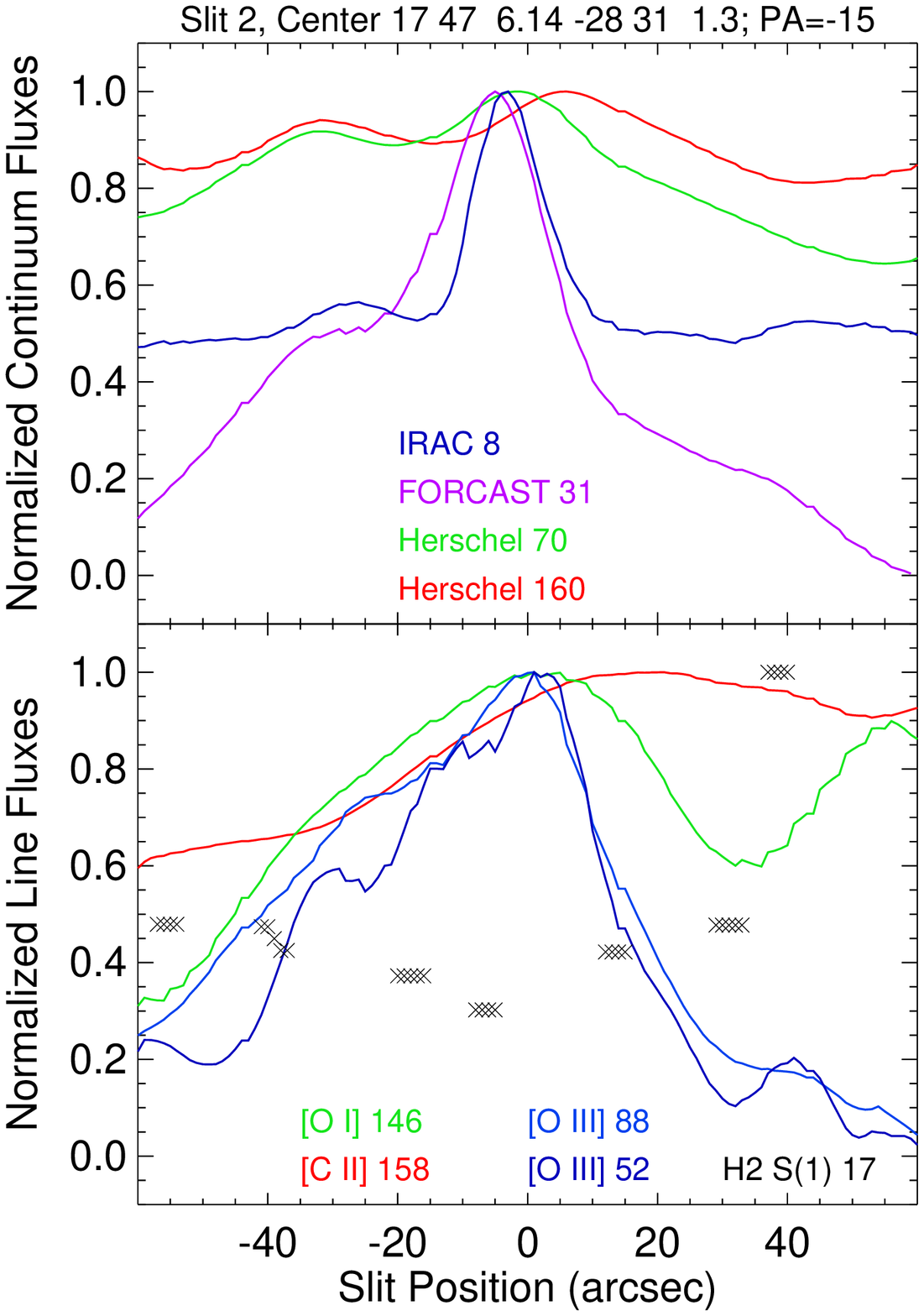}
\plottwo{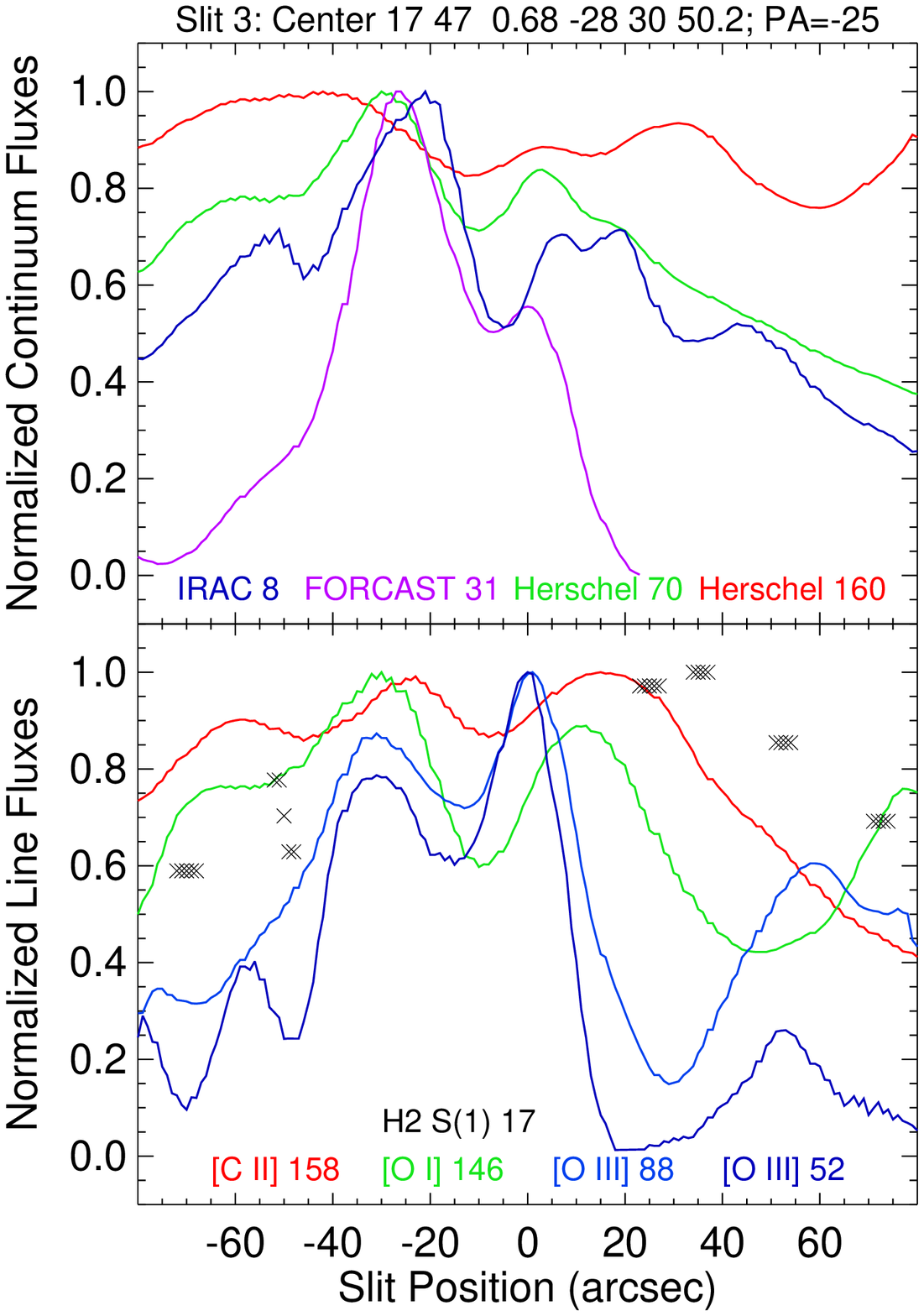}{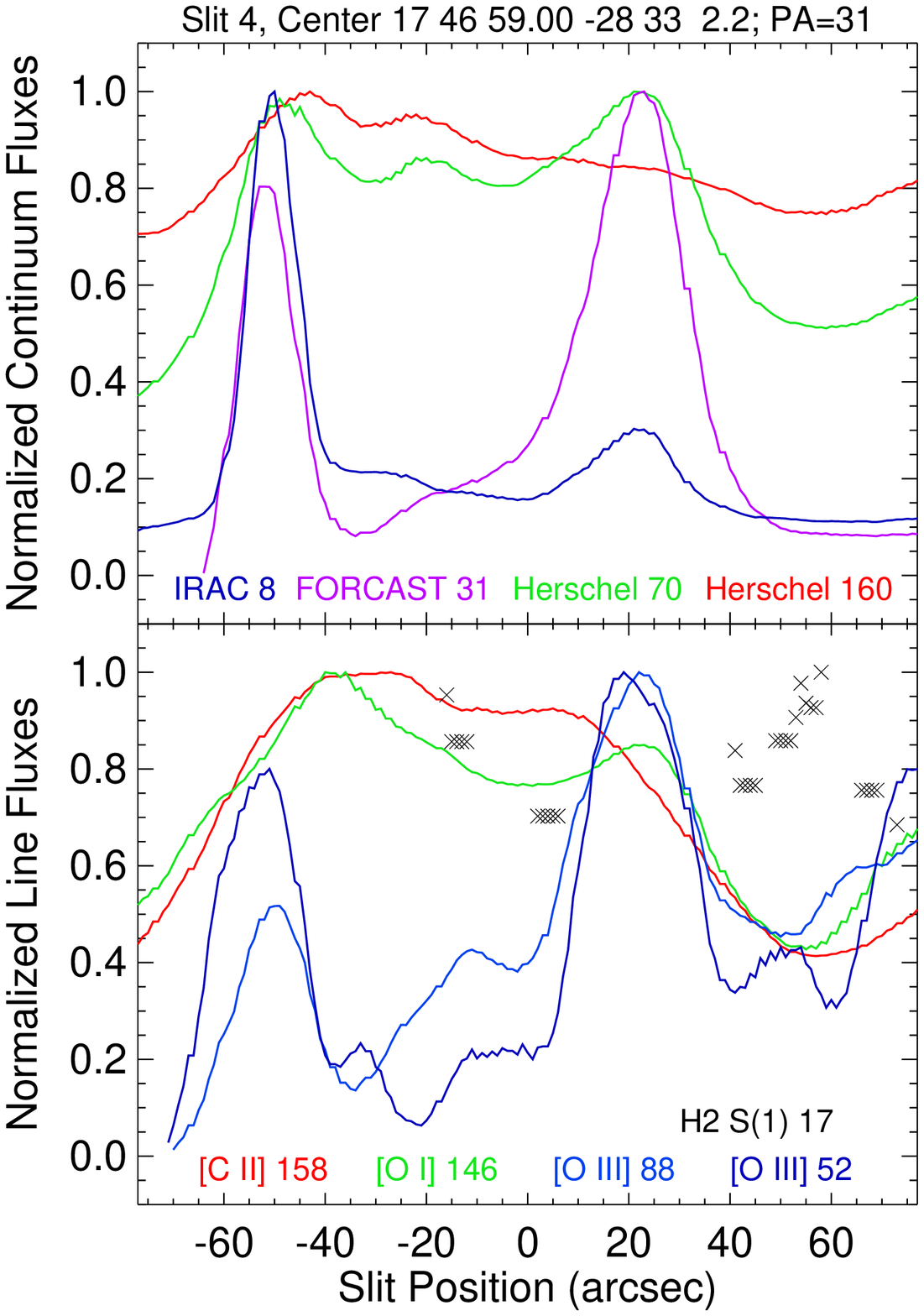}
\caption{Continuum and line profiles seen along the slits marked in Figure~3.
Coordinates of the slit centers and slit position angles (PA) are given at the tops of each panel; 
negative slit positions have more southerly declinations. 
In each group, the upper panel shows the continuum smoothed to 9\arcsec, 
the approximate resolution of SOFIA at 88~\micron, 
and the lower panel shows measurements of the four lines observed with FIFI-LS: 
[\ion{O}{3}] 52 and 88~\micron, [\ion{O}{1}] 146~\micron, and [\ion{C}{2}] 158~\micron.
Where available, {\it Spitzer} IRS measurements of the H$_2$ S(1) 17~\micron\ line 
(Simpson 2018) are marked with black X's.
}
\end{figure*}

The [\ion{C}{2}] 158~\micron\ line is very bright and was easily detected at all locations observed. 
The [\ion{O}{1}] 146~\micron\ line is less bright, with several places on the sky 
where the signal/noise was less than five even after the input data cubes 
were spatially smoothed by a factor of seven pixels prior to measuring the line intensities.
These pixels are omitted in Figure~2.

Composite images of the lines and continua are shown in Figure~3 and 
cuts through interesting regions are shown in Figure~4. 
Data plotted in the virtual slits of Figure~4 are integrated over 9\arcsec, 
the spatial resolution of SOFIA at 88~\micron. 

The immediately striking thing about Figure~3 is how different the images 
in the various continuum wavelengths and lines appear on the sky.
The continuum image best correlated with the ionized gas seen in the 
8.4~GHz radio image of Mehringer et al. (1992) 
is the {\it Herschel} PACS 70~\micron\ image (green) originating in warm dust.
The ionized gas is also bright in two to four positions 
in the IRAC 8~\micron\ image (blue),
showing that these regions have significant numbers of the FUV photons 
accompanying the extreme-ultra-violet (EUV) photons that ionize the \ion{H}{2} regions components.

On the other hand, there is little correlation with the cold dust seen 
in the 160~\micron\ {\it Herschel} PACS image (red).
The 160~\micron\ emission is strongest along the low Galactic-latitude edge 
of the Dust Ridge and the southern end of Sgr~B2, 
although there are also regions emitting strongly at 160~\micron\ in Sgr~B1 at around 
R.A. 17$^{\rm h}$47$^{\rm m}$0$^{\rm s}$ decl. $-$28\degr~33\arcmin~0\arcsec\ 
and R.A. 17$^{\rm h}$47$^{\rm m}$2\fs5 decl. $-$28\degr~31\arcmin~30\arcsec. 
Thus we included it in our analysis. 

The bottom panel of Figure~3 shows lines measured with FIFI-LS: 
[\ion{O}{3}] 88, [\ion{O}{1}] 146, and [\ion{C}{2}] 158~\micron.
The intensities of the [\ion{O}{3}] lines do not correlate well with the radio emission, 
showing that the hottest stars are found only in the eastern part of Sgr~B1, 
although there must be a sizeable number of photons from stars of cooler effective
temperatures ($T_{\rm eff}$) exciting the south-western part of Sgr~B1.
The PDR lines, [\ion{O}{1}] 146 and [\ion{C}{2}] 158~\micron, do not correlate well 
either with each other or with the continuum. 
This is seen best in the four artificial slits (Figure~4) placed on Figure~3.

Slit 1 (Figure~4, top left) is the best example of an edge-on PDR in the dataset.
The peaks of the [\ion{O}{3}] lines are well-separated from the peaks of the two 
PDR lines, and the rise to the peak of the [\ion{O}{1}] 146~\micron\ line 
at negative slit positions is much sharper than the rise in the [\ion{C}{2}] 158~\micron\ line, 
which has substantial [\ion{C}{2}] 158~\micron\ emission within the \ion{H}{2} region.
From the minimal [O I] 146~\micron\ line emission at the location of the \ion{H}{2} region, we infer that there is no background PDR or associated molecular cloud at that location.
In the continuum, the FORCAST 31~\micron\ emission is the only wavelength 
confined to the \ion{H}{2} region 
whereas the IRAC 8~\micron\ and the {\it Herschel} 70~\micron\ emission 
both arise in the \ion{H}{2} region and in the PDR. 
The cool dust emission at 160~\micron\ shows very little enhancement at the location 
of the \ion{H}{2} region --- the ends of the slit are dominated by emission from 
the cool dust of the background/foreground molecular clouds.

Slit~2 (Figure~4, top right) shows a PDR--\ion{H}{2} region face that is most likely angled to the line of sight 
since the [\ion{O}{3}] and [\ion{O}{1}] line emission 
coincide fairly well with each other and with the hot-dust emission marked 
by the IRAC 8~\micron\ and FORCAST 31~\micron\ emission.
The centroids of the [\ion{O}{3}] line emission are shifted towards negative 
slit positions 
and the centroid of the [\ion{C}{2}] emission towards positive slit position.
At even higher slit position there is a peak in the H$_2$ S(1) emission 
detected by {\it Spitzer} IRS (Simpson 2018). 
We note here that the {\it Spitzer} IRS slits have only partial coverage of Sgr~B1 
(see figure~2d of Simpson et al. 2018 for the spatial coverage of the long-low order-2 slit, 
which includes the 17~\micron\ wavelength), 
and that if the 17~\micron\ continuum was strong, the H$_2$ 17~\micron\ line 
could not be reliably measured because of low line/continuum ratios
(e.g., most of slit~1).

Slit~3 (Figure~4, bottom left) shows a complex region with multiple exciting stars 
(the [\ion{O}{3}] lines) 
and PDRs both approximately face-on (slit position $-30\arcsec$) 
and edge-on (slit positions 0 to $\sim 40\arcsec$). 
The far north end of this slit picks up emission from the Dust Ridge,
seen in both the 160~\micron\ continuum emission and in the [\ion{O}{1}] 146~\micron\ line.
The Dust Ridge is also seen in absorption in Figure~1 in the region marked by $b = 0$. 

Slit~4 (Figure~4, bottom right) shows two very distinct \ion{H}{2} regions (the [\ion{O}{3}] lines and 
the 8, 31, and 70~\micron\ continua).
The [\ion{O}{1}] and [\ion{C}{2}] lines indicate that the PDR lies between the two \ion{H}{2} regions, 
probably meaning that it is illuminated from both sides. 
The H$_2$ emission at the north end of the slit 
possibly arises in some other cloud along the line of sight.

Clearly, Sgr~B1 is a complicated region containing numerous ionizing stars 
and associated \ion{H}{2} regions. 
The PDRs on the molecular cloud edges can be face-on or edge-on, or 
anywhere intermediate.
As it appears from the earth, Sgr~B1 is a patchwork of \ion{H}{2} regions and PDRs.

\section{Discussion}

We use the software mentioned in the Introduction to model the lines observed in Sgr~B1.

\subsection{Estimates of $n$ and $G_0$ from the PDR Toolbox}

We estimated values of the hydrogen nucleus density $n$ 
and the incident flux $G_0$ at the face of each PDR, that is, for each pixel
in our observed SOFIA FIFI-LS map,
using the online PDR Toolbox\footnote{http://dustem.astro.umd.edu/}
(PDRT; Kaufman et al. 2006; Pound \& Wolfire 2008, all documentation available at this website).
(We note that Pound \& Wolfire are currently updating the PDRT to a Python application; this paper used the no-longer available `Classic' PDRT.)
To use the PDRT, one gives as input at least three observed line 
or continuum intensities and their errors, 
from which the PDRT least-squares program can form two or more ratios. 
These ratios are compared to the various ratios computed by Kaufman et al. 
(1999, 2006) and available on the PDRT website. 
In the PDRT, reduced $\chi^2$ deviations are calculated for every point 
in the $n,G_0$ plane (resolution 0.25 dex) 
and the minimum $\chi^2$ is used to estimate the appropriate  
values of $n$ and $G_0$ for the input line intensities. 
For the Classic PDRT, this could be done either one at a time online or 
through Python or Interactive Data Language 
(IDL) scripts that accessed the server running the PDRT.

In this analysis, our three input intensities were our measurements of the [\ion{O}{1}] 146 and [\ion{C}{2}] 158~\micron\ lines  
and the estimated integrated FIR continuum intensity. 
For the latter, we used the {\it Herschel} PACS and SPIRE 
70 -- 500~\micron\ intensity maps of Molinari et al. (2016) 
and the SOFIA FORCAST 31~\micron\ map of A. Cotera et al. (in preparation).
Our procedure was as follows: 
we first computed color temperatures between the observed intensities at 70 and 160~\micron. 
We then estimated a FIR spectral energy distribution (SED) for each pixel  
from 20~\micron\ to 160~\micron, employing a black body with the computed color temperatures, and normalized the blackbody intensities to the observed intensities at 70 and 160~\micron. For the longer wavelengths, we interpolated (log intensity, log wavelength) the observed intensities between 160 (PACS) and 500~\micron\ (SPIRE 250, 350, and 500~\micron) and extrapolated the intensity from 500~\micron\ to 1~mm.
Finally, we integrated over the estimated SED, normalizing the SED 
to the 70~\micron\ intensities 
that would be observed by a telescope that chopped with the same 
chopper throw and chop angle as we used with SOFIA FIFI-LS.
The color temperatures ranged from 28~K to 70~K. 
Since the 31~\micron\ map covers a smaller area on the sky than our 
observations, we added in the additional flux from 31~\micron\  
to 70~\micron\ to the estimated flux as a secondary effect. 
First we corrected the 31/70~\micron\ ratio for interstellar extinction, 
using the GC extinction map of Simpson (2018), 
the ratio of extinction coefficients at 31~\micron\ to 9.6~\micron\ 
of Chiar \& Tielens (2006), 
and the assumption that the extinction coefficients decrease from 31~\micron\ 
to 70~\micron\ as the wavelength to the minus 2 power. 
Next, as above, we computed the color temperatures between 31 and 70~\micron\ 
and the resulting normalized black-body intensities between 20 and 70~\micron. 
If the observed 31~\micron\ intensity was higher than the estimated 31~\micron\ intensity from our previously computed SEDs, we replaced that part of the estimated SED from 20 -- 70~\micron\ with the 20 -- 70~\micron\ SED with the higher intensity.
The regions with the extra intensity from the 31~\micron\ maps 
lie almost exclusively in those regions of Figure~3(a) that 
contain noticeable flux from the {\it Spitzer} IRAC 8~\micron\ image.
The increase in FIR flux due to including the 31~\micron\ FORCAST measurement is $< 20\%$ for the parts of our maps with good signal/noise ratios in the [\ion{O}{1}] 146~\micron\ line. 
The three most prominent positions with increased flux are the compact H II regions at 
R.A. (J2000) 17$^{\rm h}$46$^{\rm m}$57\fs0 decl. (J2000) $-$28\degr~33\arcmin~46\arcsec\ and 
17$^{\rm h}$47$^{\rm m}$12\fs5 $-$28\degr~31\arcmin~20\arcsec,
and the YSO at 
17$^{\rm h}$46$^{\rm m}$53\fs3 $-$28\degr~32\arcmin~01\arcsec\ 
(An et al. 2017).
For all three, it is quite reasonable to think that there
are multiple components with higher dust temperatures that should add to the FIR flux
at the shorter wavelengths.

The SOFIA FORCAST measurements of the dust in Sgr~B1 will be discussed in more detail by A. Cotera et al. (in preparation).


Using the observed [\ion{O}{1}] 146~\micron\ and [\ion{C}{2}] 158~\micron\ intensities 
and the FIR continuum described above
(with the factor of two divisor for the assumption that these GC clouds are illuminated on all sides, Kaufman et al. 1999), 
we used the PDRT IDL script that accessed the Classic PDRT 
to estimate $n$ and $G_0$ 
for every position with adequate signal/noise ratios in both lines. 
Uncertainties of 10\%\ were assumed for the FIR continuum 
and an additional 5\%\ was added in quadrature to the line errors 
for systematic uncertainties in the line measurement procedure 
(Simpson et al. 2018).
Calibration uncertainties were not included. 
Although this procedure is straight forward, 
some of the resulting $n,G_0$ are clearly not correct, 
as we describe next.

\begin{figure}
\centering
\plotone{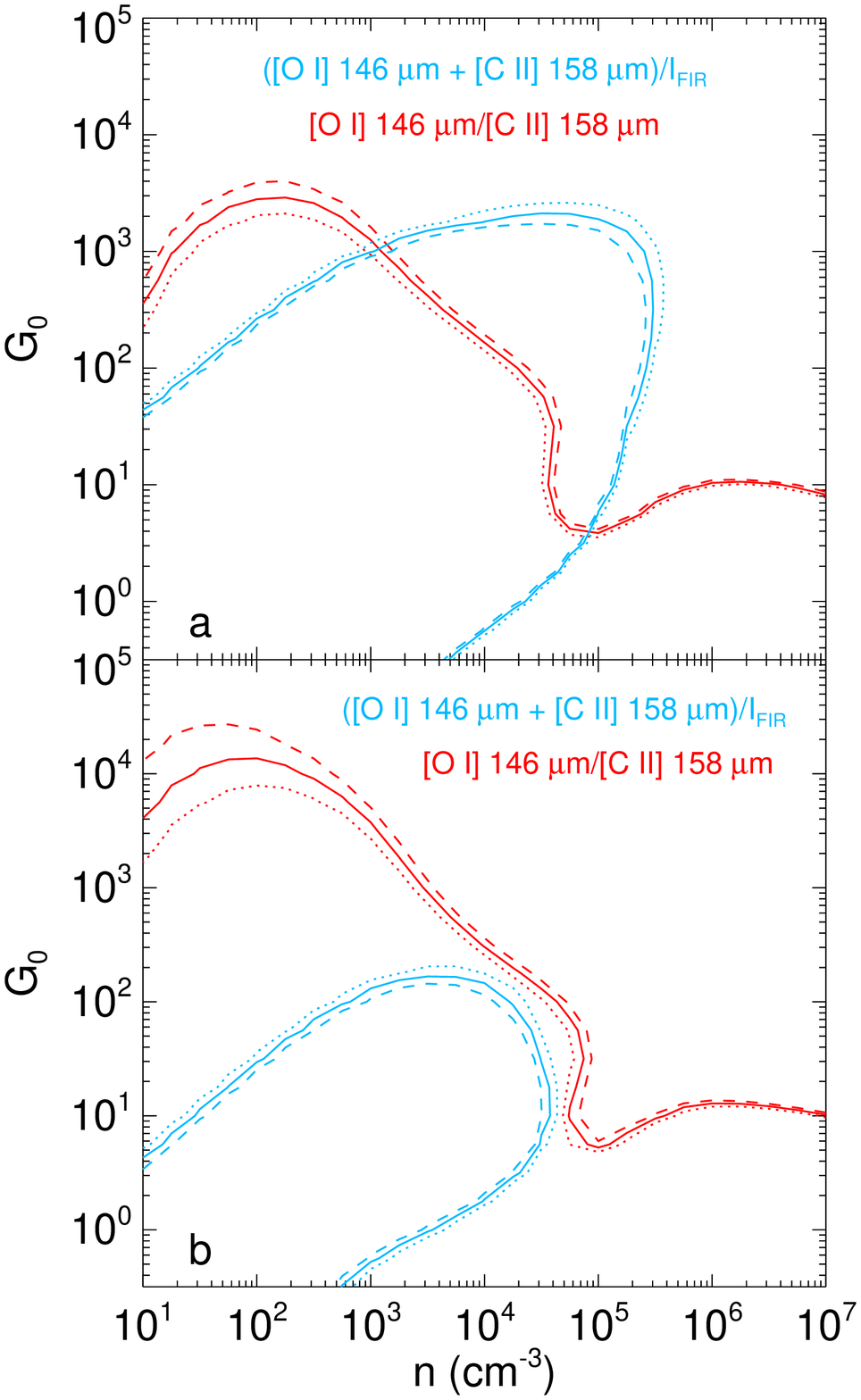}
\caption{
Loci of sample values of the [\ion{O}{1}] 146/[\ion{C}{2}] 158 \micron\ ratio 
and the ([\ion{O}{1}] 146 \micron\ + [\ion{C}{2}] 158 \micron)/$I_{\rm FIR}$ intensity ratio  
in the $n,G_0$ plane as found in the contour plots of these ratios 
by Kaufman et al. (1999, 2006) and downloaded from the PDRT.
In both panels, 
the solid blue line is the locus of the ([\ion{O}{1}] 146 \micron\ + [\ion{C}{2}] 158 \micron)/$I_{\rm FIR}$ intensity ratio  
and the solid red line is the locus of the [\ion{O}{1}] 146/[\ion{C}{2}] 158 \micron\ ratio. 
The dashed lines are the sample values plus uncertainties of 10\%\ 
and the dotted lines are the input values minus uncertainties of 10\%.
(a) The ratio represented by the blue line has a value of 0.00171 
and the ratio represented by the red line has a value of 0.1097. 
At the location of this sample, 
R.A. 17$^{\rm h}$47$^{\rm m}$01\fs0, decl. $-$28\degr~30\arcmin~47\arcsec, 
both the [\ion{O}{1}] and [\ion{C}{2}] lines are very bright and the measured 
uncertainty in their ratio is only 1\%; the uncertainty was increased 
for this plot to show the direction the loci would move if the ratios were increased or decreased.
(b) Ratios at R.A. 17$^{\rm h}$46$^{\rm m}$54\fs3, decl. $-$28\degr~30\arcmin~30\arcsec\ in the Dust Ridge. 
The ratio represented by the blue line has a value of 0.00732  
and the ratio represented by the red line has a value of 0.1587. 
The minimum $\chi^2$ ($ = 1.88$) found by the PDRT has $n = 17800$~cm$^{-3}$ and $G_0 = 178$.}

\end{figure}

Figure 5 shows examples from the PDRT for two of the positions in our SOFIA FIFI-LS map.
In panel (a), we see that the loci of the two ratios cross in two positions 
in the $n,G_0$ plane --- once near $n \sim 10^5$ cm$^{-3}$ and $G_0 \sim 3$, 
and a second near $n \sim 10^3$ cm$^{-3}$ and $G_0 \sim 10^3$.
For many map positions, there is a third crossing near $n \sim 10$ cm$^{-3}$ and $G_0 \sim 10^2$.
The minimum of the reduced $\chi^2$ is usually at one of these crossings 
for the three line/continuum intensities (two ratios) that we are using in this paper. 
Because Sgr~B1 is a very bright PDR/\ion{H}{2} region at all locations, 
one would expect the crossing with the highest $G_0$ to have the minimum $\chi^2$. 
It often happens, however, that the minimum $\chi^2$ is from a crossing with 
a computed $n,G_0$ position that has $G_0$ as low as 10.  
For these cases, we reconsider our solution
(note that particular crossings are not always statistically significant).

The PDRT considers only ratios in its least-squares
minimization because its models are all face-on but the
observed PDR’s are usually seen at some angle. Although
the solution for $n, G_0$ for each position was based
on ratios of observed intensities, for each pixel we do
have the additional information of the observed absolute
intensity, which we could compare to the predicted
intensities for the $n, G_0$ at that location. 
To refine the PDRT solution, we examined the ratios of
the observed line intensities divided by the predicted intensities 
(using either [\ion{O}{1}] or [\ion{C}{2}] gave the same results). 

When we plotted the ratio of observed/predicted intensity versus $G_0$ or $n$,
we found that approximately 11\%\ of the total number of pixels 
were distinctly separated from the rest of the pixels on the plots --- 
they had high observed/predicted ratios and low $G_0$ on the $G_0$ plot 
and high observed/predicted ratios and high $n$ on the $n$ plot. 
The dividing lines were observed/predicted [\ion{C}{2}] ratio $\gtrsim 125$, 
$G_0 < 10$, and $n \gtrsim 2 \times 10^4$~cm$^{-3}$. 
Upon further investigation, we found that 
these were the pixels where the PDRT-estimated minimum $\chi^2$ lay at the 
$n \sim 10^5$ cm$^{-3}$ and $G_0 \sim 3$ crossing point in Figure 5
(output from the PDRT includes a FITS file of the $\chi^2$ computed for all $n,G_0$). 
For these pixels we replaced the $n,G_0$ found at the minimum $\chi^2$ with the $n,G_0$ found at the next larger $\chi^2$ in the $n,G_0$ plane, 
that is, at one of the other crossing points. 
All but 0.4\% of the pixels had $\chi^2$ increase by less than a difference $\delta \chi^2 < 7.4$, 
and these pixels lie almost entirely on the edges of the holes in the maps 
caused by low signal/noise in the [\ion{O}{1}] 146~\micron\ line measurements. 
Consequently, we removed from our $n,G_0$ maps all those pixels with $\chi^2 \gtrsim 7.6$ 
so that all remaining pixels now have $\chi^2 < 7.6$. 

Figure 5(b) shows an example with no crossings. 
Pixels with this pattern of tangent or no crossings all have the [\ion{O}{1}] 146/[\ion{C}{2}] 158 ratio $> 0.145$ and the ([\ion{C}{2}] 158 + [\ion{O}{1}] 146)/$I_{\rm FIR}$ ratio $ > 0.004$ (i.e., exceptionally high [\ion{O}{1}] 146~\micron\ fluxes, see Figure~3b). 
These approximately 100 pixels are all found in the Dust Ridge region NW of Sgr~B1. Where the two lines are close to tangent, $\chi^2$ is very small, 
but for larger separations of the two lines (larger [\ion{O}{1}] 146/[\ion{C}{2}] 158 ratio or larger ([\ion{C}{2}] 158 + [\ion{O}{1}] 146)/$I_{\rm FIR}$ ratio), $\chi^2$ is larger.
Those pixels with $\chi^2 > 7.6$ were set to zero.
The lines of sight represented by these pixels clearly violate some of the assumptions of the PDRT; 
a possible reason is that they contain multiple PDR components, and thus some caution should be taken regarding their PDRT results. 

\begin{figure*}
\centering
\plottwo{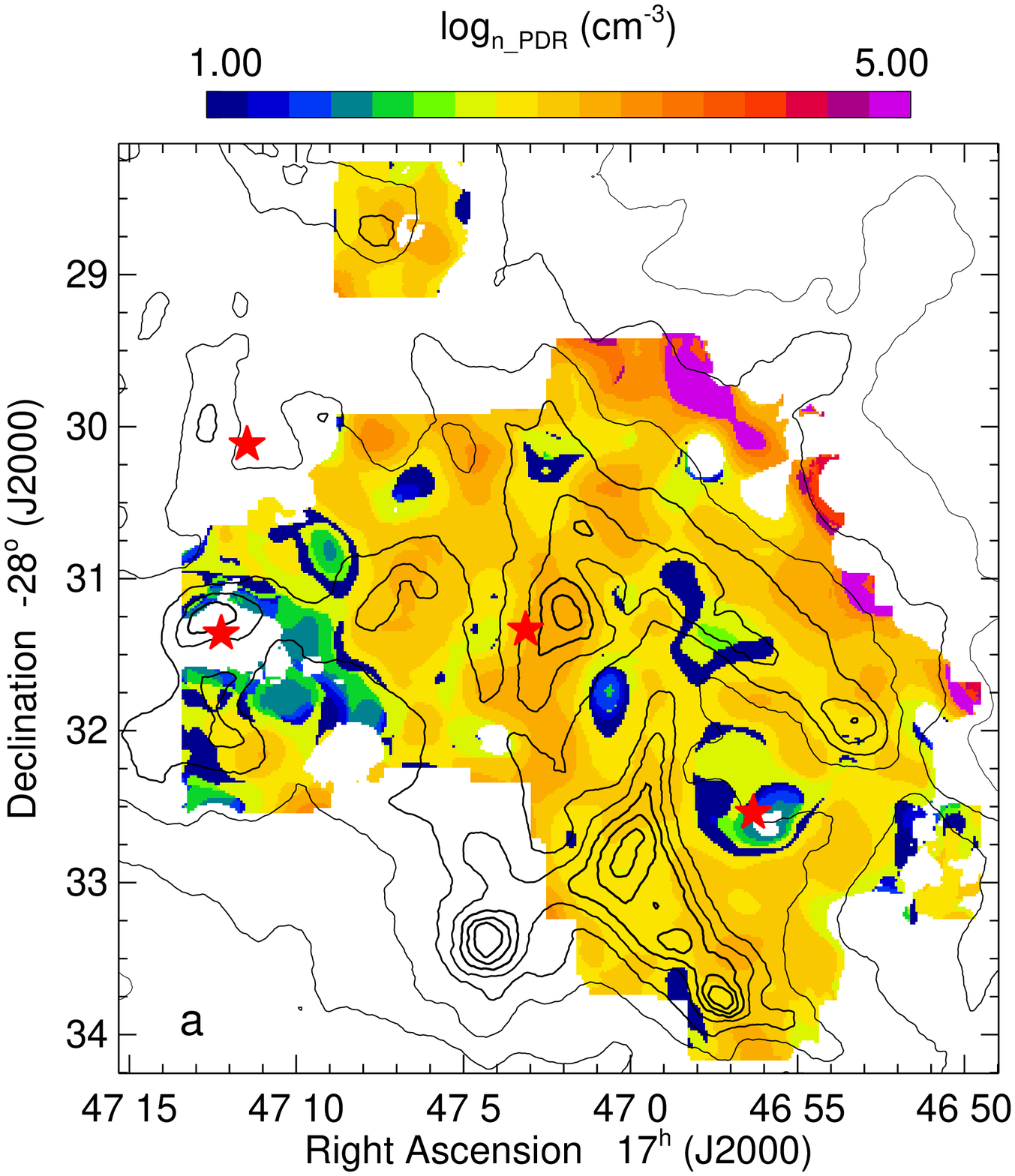}{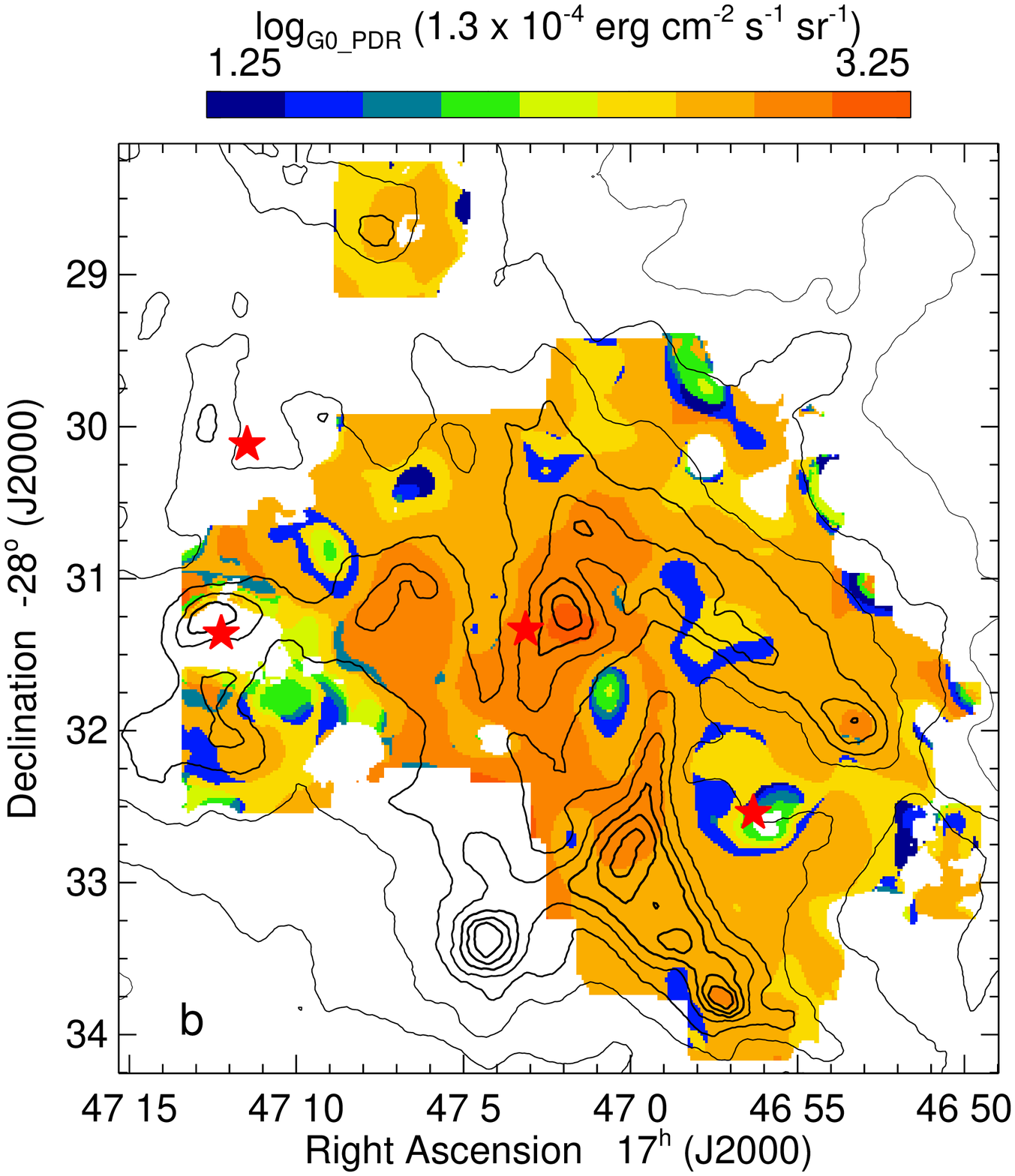}
\caption{Estimates of the hydrogen nucleus density $n$ and the incident FUV intensity $G_0$ (see text).
The black contours are the continuum intensities seen in the 70~\micron\ {\it Herschel} PACS image (Molinari et al. 2016).
The red stars are the O supergiant and WR stars identified by Mauerhan et al. (2010) that were plotted in Figure~1.
}
\end{figure*}

\begin{figure}
\plotone{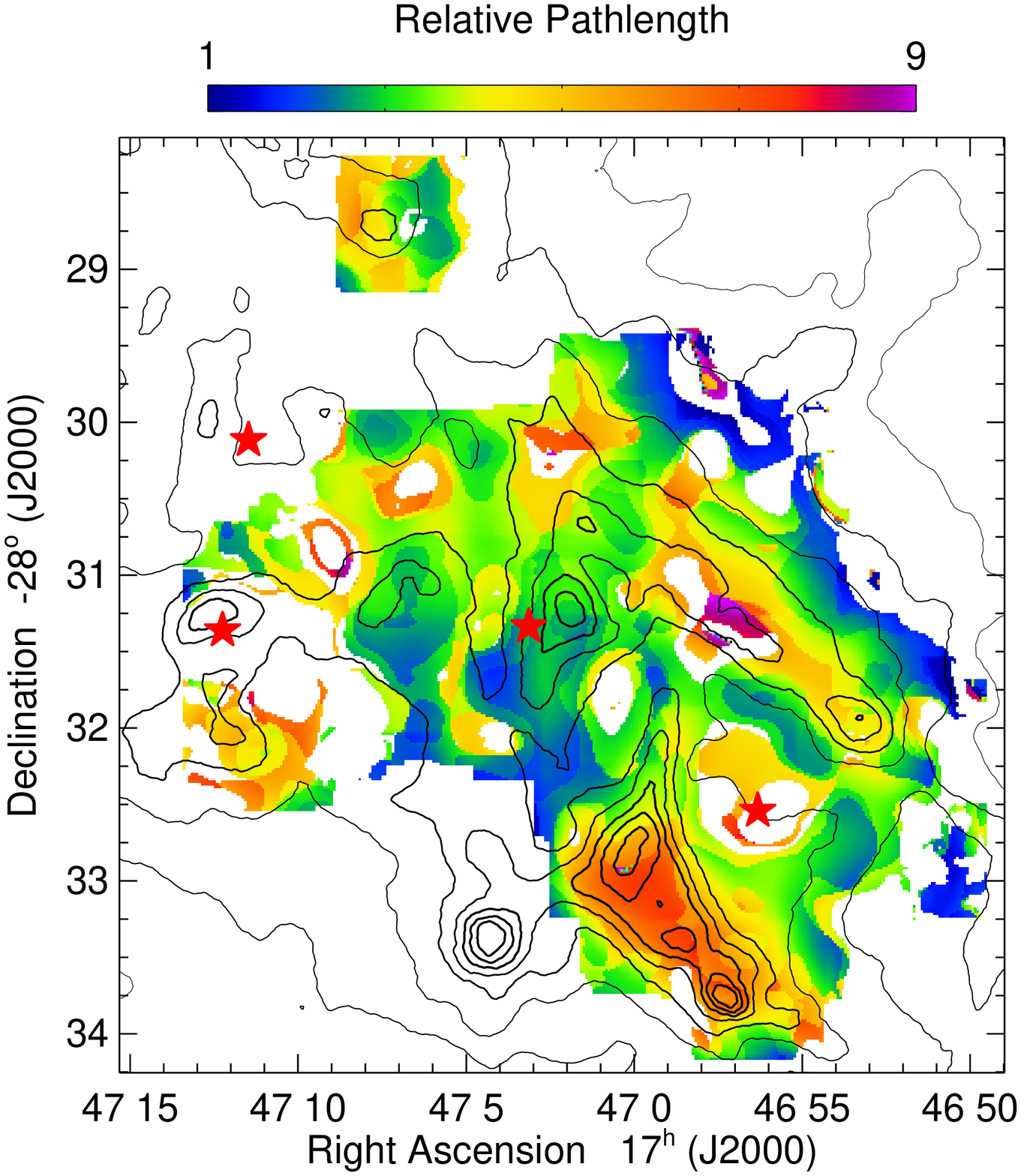}
\caption{Possible relative pathlengths of the emitting [\ion{C}{2}] 158~\micron\ line.
These are defined as the observed [\ion{C}{2}] 158 intensity divided by the intensity of the [\ion{C}{2}] 158~\micron\ line predicted for the values of $n$ and $G_0$ in Figures~6(a) and (b).
Areas of very low $n$ in Figure~6(a) are omitted because they probably arise from long paths through the diffuse ISM along the line of sight to Sgr~B1. 
The black contours are the continuum intensities seen in the 70~\micron\ {\it Herschel} PACS image (Molinari et al. 2016) and the red stars are the O supergiant and WR stars identified by Mauerhan et al. (2010).
}
\end{figure}

The results are shown in Figure~6. 
The blank holes in the figures are the locations where there was inadequate 
signal/noise in the [\ion{O}{1}] 146~\micron\ line (Figure~2a).
FITS files of $n$ and $G_0$ are included in the paper.

In Figure~6 we see that the lowest density regions also have low $G_0$; 
these $n,G_0$ solutions are the line ratio crossings at $n \sim 10$~cm$^{-3}$ of Figure~5, which occur when the [\ion{O}{1}]/[\ion{C}{2}] ratio is particularly low. 
In Figure~2(a), the low [\ion{O}{1}] 146~\micron\ intensity regions 
(and low density regions in Figure~6a, plotted in blue and green) 
appear to be located in the suggested wind-blown bubbles, such as at 
R.A. 17$^{\rm h}$46$^{\rm m}$56$^{\rm s}$ decl. $-$28\degr~32\arcmin~30\arcsec, 
where the line and continuum emission at all wavelengths are weak (Figures 1 to 3).
We note that in these regions, the ratio $n/G_0$ is less than 1. Kaufman et al. (1999) remark that in this regime, radiation pressure would drive the grains through the gas at velocities greater than the assumed average turbulent velocity of the gas. 
We do not feel this is sufficient justification to force a Figure~5 crossing at higher $n$ and $G_0$.

The relatively few pixels with the highest densities occur in the highest Galactic latitudes, where the Dust Ridge becomes apparent in the continuum emission at the longest wavelengths. 
For these pixels, the [\ion{O}{1}] 146~\micron\ line is strong and $G_0$ is low, the condition described in Figure~5(b). 

On the other hand, the locations of the highest $G_0$ in Figure~6(b) are well correlated with the peaks of the 8~\micron\ emission seen in Figure~1 and the 8~\micron\ blue image of Figure~3(a). 
We note that these are also the regions of high IRAC 8.0/5.8~\micron\ ratios that Arendt et al. (2008) suggested are due to high incident radiation field intensity heating the dust grains rather than high PAH abundance. 
The MIR colors and the dust temperatures of Sgr~B1 will be discussed in more detail by A. Cotera (in preparation).

Interestingly, we see more correlation of $G_0$ in Figure~6(b) with the contours 
of the 70~\micron\ {\it Herschel} PACS emission (Molinari et al. 2016) 
than we do with $n$ in Figure~6(a).  
This can occur when increased 70~\micron\ emission, as seen in a contour map, 
is the result of a longer pathlength seen observing through the side of a shell 
but the exciting source of the gas in the shell is close to the shell edge. 
Figure~7 shows the pathlengths estimated by dividing the observed [\ion{C}{2}] intensities by the intensities predicted by the PDR models of Kaufman et al. (1999, 2006) for the $n$ and $G_0$ plotted in Figure~6. Interestingly, although the high intensities seen in the Herschel 70~\micron\ images and the line intensities of Figure~1 surrounding the bubble at 
R.A. 17$^{\rm h}$46$^{\rm m}$56$^{\rm s}$ decl. $-$28\degr~32\arcmin~30\arcsec\ 
do indeed show longer pathlengths for the bubble rim, 
the location of the peak $G_0$ at 
R.A. 17$^{\rm h}$47$^{\rm m}$2$^{\rm s}$ decl. $-$28\degr~31\arcmin~15\arcsec\ 
does not have an exceptionally long pathlength. 
We infer that the gas in this region is truly more compact and that its ionizing star must be relatively close.

We can compare the densities estimated for the ionized gas in the Sgr~B1 \ion{H}{2} region 
and the density in the PDR as computed by the PDRT. 
There is no correlation between the electron density $N_e$ computed 
from the [\ion{O}{3}] lines of Simpson et al. (2018) 
and the hydrogen nucleus density $n$ from the PDRT (correlation coefficient $R = 0.055$).
Using only those pixels that have good measurements in both the electron density map 
of Simpson et al. (2018) and in the $n$ map of Figure~6(a), 
we find that the medians of these two measurements (\ion{H}{2} region and PDR) 
are 240 and 560~cm$^{-3}$ 
and the averages are 310 and 940~cm$^{-3}$, respectively. 

The lack of a strong pixel-to-pixel correlation in the densities 
is another indication that most of the PDRs that we see in Sgr~B1 
are not face-on but are at least somewhat edge-on. 
Moreover, there is no over-riding structure in the density 
that would be indicative of any underlying molecular cloud structure --- 
no centralized cloud core nor any hierarchical structure 
with numerous cloud clumps. 
In particular, there are no pixels with the high levels of $G_0$ ($\gtrsim 10^{3.5}$) 
that would come from a cluster containing nearby stars; 
the numerous OB stars that are required to produce the $3 \times 10^{50}$~s$^{-1}$ ionizing photons 
can produce the observed fairly uniform $G_0$ if they are also widely spread out.  
We conclude that our estimates of the PDR $n$ and $G_0$ are 
yet additional indicators of substantial evolution and dispersal 
with no current clustered massive star formation.

\subsection{Edge-on PDR Models Computed with Cloudy}

\input tab2.tex

\begin{figure}
\centering
\plotone{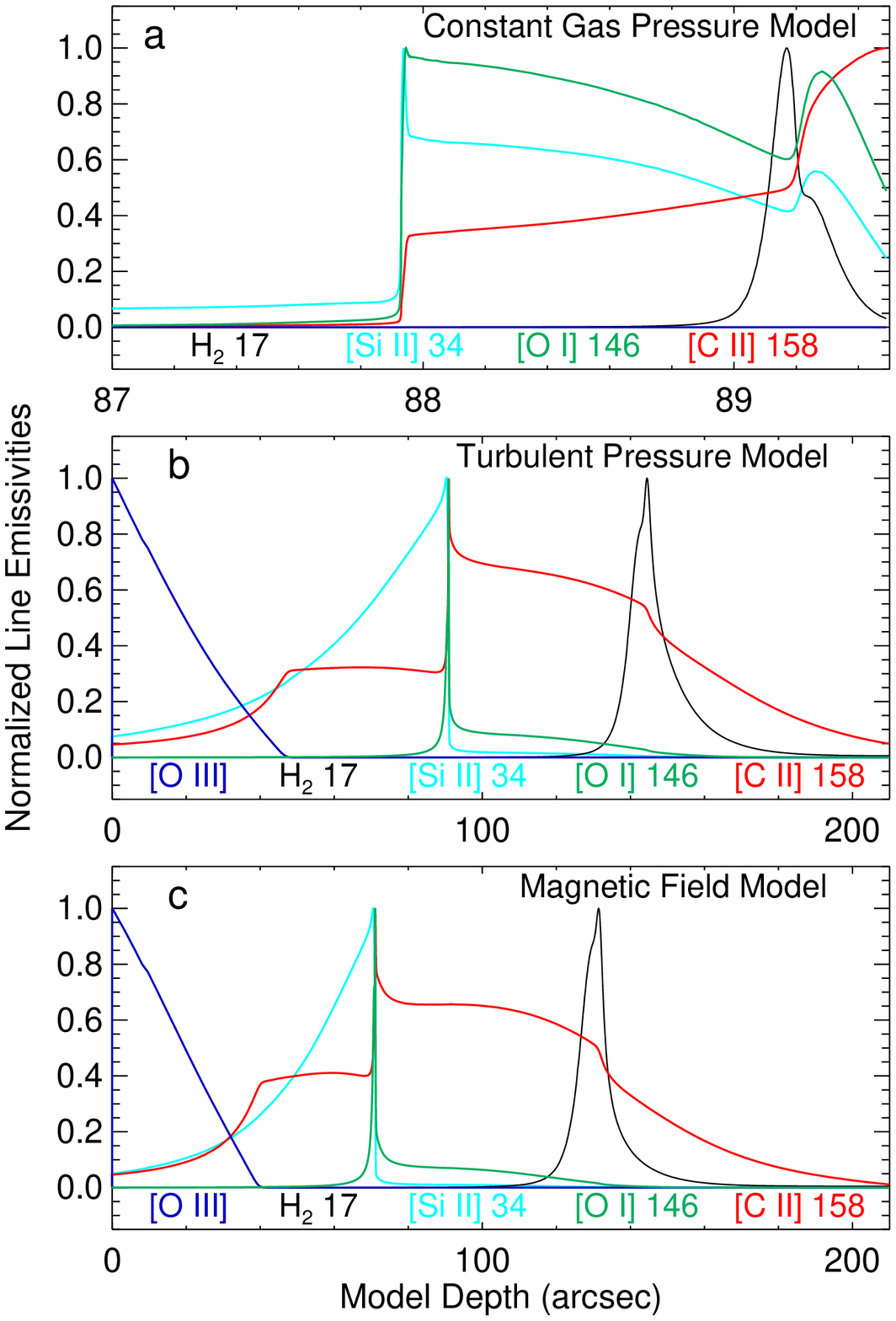}
\caption{
Normalized volume emissivities from Cloudy models 
showing the effects of adding pressure terms to constant pressure models. 
The different lines are plotted in different colors as shown. 
The abscissa (arcsec) is scaled to an assumed distance of 8~kpc for Sgr~B1.
The face of the PDR occurs at the peak of the [\ion{O}{1}] line emissivity.  
See the text and Table 2 for the model parameters. 
(a) Constant gas pressure with no added pressure. Note that the inner edge of the \ion{H}{2} region is at Model Depth $= 0\arcsec$ but only the thin PDR is plotted.  
(b) Constant gas plus turbulent pressure.  
(c) Constant gas plus magnetic pressure.  
}
\end{figure}

We have seen that Sgr~B1 contains PDRs, both face-on and edge-on, 
with the edge-on examples showing separations of the \ion{H}{2} region and the PDR 
by sometimes over an arcminute.
However, PDR models produced assuming constant gas pressure, 
such as those of Kaufman et al. (2006), 
have quite thin PDRs separating their \ion{H}{2} regions and molecular clouds. 
The reason is that, with the very large change in gas temperature 
going from an \ion{H}{2} region (electron temperature $T_e \sim 6500$~K in the GC) 
to a molecular cloud (gas temperature sometimes as low as 10~K), 
the gas densities in the PDRs can be orders of magnitude larger than 
in the \ion{H}{2} region, 
with the result that the FUV photon density decreases 
over a very short distance compared to the size of the \ion{H}{2} region. 
Here we discuss possible reasons why the PDRs of Sgr~B1 can have 
the extended PDRs that we observe. 

The classic example of an edge-on PDR is the relatively nearby ($\leq 0.5$~kpc) Orion Nebula Bar, observed by Tielens et al. (1993). 
Their paper presents models of the PAH features and H$_2$ and CO(1--0) lines,
which are separated from each other in layers about 10\arcsec\ 
($\sim 0.2$~pc) apart. 

Pellegrini et al. (2009) and Shaw et al. (2009) also 
made models of the Orion Nebula Bar, 
using the PDR capabilities of the \ion{H}{2} region code Cloudy  
(Abel et al. 2005; Shaw et al. 2005; Ferland et al. 2017). 
They found good agreement of the observations with 
their predicted variations with distance from the exciting star 
of the Orion Nebula, $\theta^1$~C~Ori, in lines of 
H$\alpha$, [S II], H$_2$, [\ion{O}{1}], [\ion{C}{1}], 
and various molecules including CO (Pellegrini et al. 2009) 
and H$_2$ (Shaw et al. 2009). 
In particular, 
Pellegrini et al. (2007, 2009) demonstrated that the physical depth of the PDR 
and the separation of the atomic hydrogen and H$_2$ lines 
is dependent on the gas density within the PDR. 
To show this, they computed constant pressure models that included various amounts of magnetic field, cosmic rays, and turbulent pressure.   
These added pressure and/or heating components 
enabled them to produce models of M17 and the Orion Nebula Bar 
(Pellegrini et al. 2007, 2009, respectively) that reasonably match the observations.

However, the Orion Nebula Bar is at a distance of $\sim 0.5$~kpc, 
and the observed separations between the different PDR lines 
of order 10\arcsec, regardless of the cause,  
would not be detectable at the $\sim 8$~kpc distance of Sgr~B1.   
In order to understand the PDRs in Sgr~B1 and what is needed 
to produce examples similar to what we see, we have computed 
grids of models using 
Cloudy 17.01\footnote{https://www.nublado.org} (Ferland et al. 2017).
We discuss here three PDR models that reproduce the various lines 
that have been observed in Sgr~B1; 
parameters of the models are given in Table~2. 
The \ion{H}{2} region part of the model was chosen to give good agreement 
with the Sgr~B1 results of Simpson et al. (2018) and Simpson (2018) --- 
ionization by a star of fairly low $T_{\rm eff}$, 
moderately low electron density $N_e$, 
but high enough ionizing photon density (represented by the ionization parameter $U$)
that silicon and sulfur are mostly doubly ionized 
(see the discussion in Simpson 2018 on the use of the [\ion{S}{3}] 33/[\ion{Si}{2}] 34~\micron\ ratio 
to estimate the value of $U$). 
Our examples here are not any `best fit' that can be used 
to characterize Sgr~B1 but instead are designed to explore 
the geometry and physics of the region.

Cloudy has a number of options for treating a PDR excited 
by the radiation field emitted by an \ion{H}{2} region. 
In addition to specifying a varying density law (possibly from a table) 
of densities from the surface of the \ion{H}{2} region through the ionization front 
to the depths of a molecular cloud, 
Cloudy can use either a constant density (the default) 
or compute the density assuming constant pressure for all depths of the model. 
The simplest constant pressure models assume that the pressure  
is given by the ideal gas law (density $\propto 1/T$), 
with both density and temperature changing accordingly. 
However, the result is that the density in the PDR is orders of magnitude 
higher than in the \ion{H}{2} region because of the great decrease in 
gas temperature going from the \ion{H}{2} region to the depths of the PDR. 
Such PDRs are very thin, too thin to have their different lines 
separated to the extent that we observed in Sgr~B1.

Cloudy also allows for the addition of turbulent pressure or 
pressure due to a magnetic field in constant pressure models. 
We investigated both possibilities since the GC is known to 
be turbulent (e.g., Dale et al. 2019; Kruijssen et al. 2019) 
and to contain a sizeable magnetic field 
(e.g., Morris \& Serabyn 1996; Ferri\`ere 2009).
To test the effects of adding additional pressure terms to 
constant pressure models, 
we computed three models with the same input \ion{H}{2} region parameters:  
a model with constant gas pressure only (density $\propto 1/T$), 
a model with constant pressure consisting of gas pressure and turbulent pressure, 
and a model with constant pressure consisting of gas pressure and magnetic field pressure. 
We note that all three models include a term for `turbulent' line broadening in addition to the thermal line broadening in order to increase the widths of the line profiles 
(particularly the [\ion{C}{2}] 158~\micron\ line, which otherwise is quite optically thick); 
in Cloudy, it is optional to include or not include the turbulence that is used in the line width specification in the terms used in the pressure equilibrium calculation. 
The equations describing these terms are given in the Cloudy user's guide,
Hazy1.pdf (Ferland et al. 2013, 2017). 

The results are given in Table~2 and examples of 
normalized line emissivities as a function of depth from the star-facing edge 
of the \ion{H}{2} region (scaled to the $\sim 8$~kpc distance of Sgr~B1) are given in Figure~8.
Line emissivities are plotted rather than the more usual fractional ionizations 
so that the reader can visualize how these lines would appear 
when viewing an edge-on \ion{H}{2} region plus PDR. 
Only the PDR and the ionization front of the constant gas density model are plotted in Figure 8(a). 

The two methods of adding extra pressure to the PDR and hence reducing 
its density are seen to both reduce the density of the PDR and to spatially separate  
the predicted volume emissivities of the various lines plotted
(compare the \ion{H}{2} region density at the innermost zone, $n = 316.2$~cm$^{-3}$ with the PDR densities $n$ at the peak of the H$_2$ S(1) 17~\micron\ emission line). 
We note the significant differences between the profiles of the emissivities 
as a function of distance between the high-PDR-density model of Figure 8(a) 
and the lower-PDR-density models of Figures 8(b) and 8(c). 
This is a function of density and also $G_0$ ---
models with even higher $n$ and $G_0$ are shown in figure 8b of 
Tielens \& Hollenbach (1985).  

In Figure 8, the location of the center of the fairly low-excitation \ion{H}{2} region is shown 
by the normalized emissivity of the [\ion{O}{3}] lines, identical on this plot.  
The face of the PDR is indicated by the sharp peak of the [\ion{O}{1}] 146~\micron\ line.
The singly ionized lines, [\ion{C}{2}] 158 and [\ion{Si}{2}] 34~\micron, 
are usually described as PDR lines because the ionization potentials 
of the C and Si atoms, 11.3 and 8.2~eV respectively, 
are lower than the 13.6~eV ionization potential of hydrogen.
We see here that while the [\ion{C}{2}] 158~\micron\ line is mostly formed in the PDR, 
in the low-density models, 
the [\ion{Si}{2}] 34~\micron\ line is mostly formed in the \ion{H}{2} region 
(see also figures~20 and 21 of Kaufman et al. 2006, who found an abundance effect 
on the relative fractions of these lines in \ion{H}{2} regions and PDRs).
In contrast to the [\ion{C}{2}] 158~\micron\ line, the [\ion{O}{1}] 146~\micron\ line 
is formed almost entirely at the face of the PDR in the low-density models. 
The reason is that the line has a relatively high energy of its upper level 
of 326~K (Kaufman et al. 1999), 
and the gas temperature is falling very rapidly 
from the \ion{H}{2} region ionization front into the PDR.
The [\ion{O}{1}] 63~\micron\ line has more emissivity than the 146~\micron\ line 
in the cooler region behind the front 
(and hence is more susceptible to optical depth effects) 
but is otherwise similar. 
Although this part of the PDR contains substantial amounts of neutral hydrogen, 
the neutral hydrogen 21~cm line has only been measured in the GC in absorption 
(e.g., Lang et al. 2010).

Figure~4 showed cuts through Sgr~B1 in both various lines and continuum wavelengths, which can be compared to the models of Figure~8. 
In Figure~4 we see that the [\ion{O}{3}] lines from the center of the \ion{H}{2} region 
nearest the exciting stars 
are mostly well-separated from the [\ion{O}{1}] 146~\micron\ line, 
and that both are compact 
compared to the [\ion{C}{2}] 158~\micron\ line. 
This is especially evident in Slit~1, which is the closest 
to being describable as an edge-on \ion{H}{2} region plus PDR in our field of view. 
We suggest that future studies of edge-on PDRs include the H$_2$ S(1) 17~\micron\ line to test the location of the peak H$_2$ emission relative to the PDR face. 

We conclude that the distinct separation of the various \ion{H}{2} region and PDR lines 
seen in Sgr~B1 (Figures 3 and 4) 
would not be possible in anything other than low-density PDRs. 
There are two possible inferences from this: 
(1) The gas in the Sgr~B1 \ion{H}{2} region  
is physically well-separated from the PDR, 
although this does not look likely considering how much the continuum 
components at 8~\micron\ (PDR, Figure~1) look like the ionized gas 
(VLA radio image of Mehringer et al. 1992; Figure~2). 
(2) The PDR gas is low density from significant addition 
to the thermal gas pressure of either turbulence or magnetic field, or both.
We regard this as more likely because in the previous section, 
the PDRs were also determined 
to be low density from direct comparison to the PDRs of the PDR Toolbox.

\subsection{Energy Input from Candidate Local Hot Stars}

From single-dish radio measurements of the flux from Sgr~B1, 
we have inferred total ionizing luminosities from all the stars 
of $\sim 3 \times 10^{50}$ photons s$^{-1}$ (Simpson 2018). 
Mehringer et al. (1992) estimated the numbers and spectral types 
of the stars (assumed zero-aged main sequence, ZAMS) that would be needed 
to ionize the various components observed in their VLA image.
Even though their list of components implies $\sim 18$~OB stars,
their total VLA flux compared to the single-dish measurements 
implies that the total number of exciting ZAMS stars is at least 4--5 times larger.

Here we consider the excitation by stars like the Wolf-Rayet (WR) 
and O supergiant stars identified in Figure~1.
We choose these stars because their locations on the sky very interestingly 
correspond to either high-excitation regions in the [\ion{O}{3}] maps 
of Simpson et al. (2018) 
or the possible voids in Figures 3 and 6 
that could occur as the result of the strong winds from WR stars. 
Spectral types of these four stars were estimated by Mauerhan et al. (2010) 
as O4-6I for 2MASS J17470314-2831200 
and nitrogen-sequence WR (WN) 
WN7-8h or WN8-9h for 2MASS J17471225-2831215, J17465629-2832325, 
and J17471147-2830069.

The numbers of hydrogen-ionizing photons per second $Q$ have been computed using 
various stellar atmosphere codes. 
Modern codes that assume non-LTE equilibrium for the various line excitation levels 
and strong stellar winds include 
WM-BASIC (Pauldrach et al. 2001), 
CMFGEN (Hillier \& Miller 1998), 
and the Potsdam Wolf-Rayet code (PoWR, Gr\"afener et al. 2002; Hamann \& Gr\"afener 2003; Sander et al. 2015). 
Sternberg et al. (2003) and Martins et al. (2005) tabulated 
$Q$ as functions of spectral type 
using the WM-BASIC and CMFGEN codes, respectively. 
For the O4-6I star in Sgr~B1, there is a star with a very similar K-band spectrum (\#9) 
in the analysis of OB stars in the VVV CL074 cluster by Martins et al. (2019), 
who described it as spectral type O4-6If+ with $T_{\rm eff} = 37000$~K and
log~$L/L_{\odot} = 6.01$. 
The calibration of Martins et al. (2005) for luminosity class I 
and $T_{\rm eff} = 37070$~K (spectral type O5.5I) gave the luminosity 
as log~$L/L_{\odot} = 5.82$ and the total $Q$ as $(Q_H + Q_{He}) = 10^{49.59}$~s$^{-1}$. 
Adjusting for the larger log~$L/L_{\odot}$ in VVV CL074 compared to the model in Martins et al. (2005), we estimate $(Q_H + Q_{He}) = 10^{49.78}$~s$^{-1}$.

The characteristic feature of WR stars is that their spectra are dominated 
by broad emission lines; 
as a result, the effective temperature and luminosity parameters of such stars 
are very difficult to determine. 
Recently, Todt et al. (2015) have updated their grid of WN star models 
and made them available on the PoWR web site\footnote{http://www.astro.physik.uni-potsdam.de/PoWR.html}.
Models that have similar K-band spectra to the WN7-8h and WN8-9h spectra 
observed by Mauerhan et al. (2010) are the PoWR WNL-H20 models 05-11 and 04-10, respectively. 
For luminosities log~$L/L_{\odot} = 5.30$, these models have log~$(Q_H + Q_{He}) = 48.96$ and 48.84, respectively.
However, it is now thought that the late-type WN stars are more luminous 
than the log~$L/L_{\odot} = 5.30$ that was input to the PoWR code --- 
using distances from Gaia DR2, Hamann et al. (2019) 
estimated luminosities of the late-type WN stars ranging from $10^{4.9}$ to $10^{6.5}$~$L_{\odot}$ 
with the median $10^{5.8}~L_{\odot}$ for the more luminous WN stars with hydrogen lines 
(like the three WR stars in Sgr~B1).
In addition, Rate \& Crowther (2020) also used the Gaia DR2 distances 
to estimate absolute V magnitudes for all Galactic WR stars bright enough to be measured by Gaia DR2. 
Guessing a bolometric correction of $-3.5$~mag for $T_{\rm eff} = 37000$~K stars (e.g., Martins et al. 2005), 
we find similar luminosities ranging from $10^{5.1}$ to $10^{6.1}~L_{\odot}$.
Correcting the ionizing luminosities for the higher observed luminosities $\sim 10^{5.8} L_{\odot}$,
we now estimate combined $(Q_H + Q_{He}) \sim 10^{49.4}$~s$^{-1}$.

Thus the four spectrally-classified stars in Figure~1 could produce as much as $1.4 \times 10^{50}$~s$^{-1}$ ionizing photons. 
Only twice this number of ionizing photons are needed to power Sgr~B1. 
We conclude, as did Simpson et al. (2018), 
that the ionizing stars (such as these OB and WR stars) 
could come from some star formation episode of several million years previous 
and that these stars are currently dispersing through the Sgr~B cloud 
and ionizing Sgr~B1. 

Finally, we consider whether these luminous stars can produce the 
estimated $G_0$ that is found in the Sgr~B1 PDRs. 
A late O supergiant with luminosity $10^{5.8}~L_{\odot}$ modeled with WM-BASIC 
from the Binary Population and Spectral Synthesis (BPASS) project (Eldridge et al. 2017; Table~2) 
would have $G_0 \sim 1.4 \times 10^3$ at a PDR face 2.3~pc from the star, 
corresponding to a distance of 60\arcsec\ at an assumed 8~kpc distance. 
Somewhat hotter supergiants ($T_{\rm eff} = 34600$~K and 37200~K) from BPASS 
of the same luminosities 
would have slightly lower $G_0 \sim 1.2 \times 10^3$ at a distance of 2.3~pc. 
These values are in excellent agreement with the ranges of $G_0$ 
seen in Figure~6(b). 
We conclude that single luminous stars, 
separated from the PDR gas and dust by as much as several pc, 
can easily explain the dispersed regions of higher excitation 
or apparent PDR faces in Sgr~B1. 

\section{Summary and Conclusions}

We present SOFIA FIFI-LS maps of the GC \ion{H}{2} region Sgr~B1 in the PDR lines of [\ion{O}{1}] 146 and [\ion{C}{2}] 158~\micron.
These are complementary to our [\ion{O}{3}] 52 and 88~\micron\ line intensity maps that we published in Simpson et al. (2018).

We plot 3-color images of the 8, 70, and 160~\micron\ continuum intensities 
and the [\ion{O}{3}] 88, [\ion{O}{1}] 146, and [\ion{C}{2}] 158~\micron\ line intensities 
(blue, green, and red, respectively) in Figure~3. 
Whereas there is substantial agreement between the 8 and 70~\micron\ continuum images, 
the line images have much less agreement with each other, 
neither the highly ionized [\ion{O}{3}] lines with the [\ion{O}{1}] and [\ion{C}{2}] PDR lines 
nor, in detail, the PDR lines with each other. 
Line cuts through both composite images are plotted in Figure~4. 
In these artificial slits we see that the lack of agreement is due 
to the \ion{H}{2} region/PDR interface being sometimes face-on and 
sometimes edge-on, or anything in between, with no coherent structure.

We used the PDR Toolbox (Kaufman et al. 2006; Pound \& Wolfire 2008) 
to estimate the hydrogen nucleus density $n$ and the FUV intensity $G_0$ 
at the face of the PDRs for all pixels with good intensities 
in the [\ion{O}{1}] 146 and [\ion{C}{2}] 158~\micron\ lines
and estimated total FIR continuum intensities $I_{\rm FIR}$. 
The median and average PDR densities are 560 and 940~cm$^{-3}$, respectively; 
in comparison, the median and average electron densities $N_e$ are 
240 and 310~cm$^{-3}$, respectively, as measured from the ratio 
of the 52 and 88~\micron\ [\ion{O}{3}] lines (Simpson et al. 2018) 
for those pixels that have good measurements in common. 
Plotted in Figure~6, both $n$ and $G_0$ are seen to have a fairly uniform structure 
across most of the Sgr~B1 region, with no high-density clumps  
or regions of very high $G_0$ that could indicate a cluster 
of massive stars. 
In fact, most of the source has $G_0$ between 1000 and 1778
($G_0$ is the ratio of the incident FUV intensity to that of the local interstellar intensity, $1.3 \times 10^{-4}$ erg~cm$^{-2}$~s$^{-1}$~sr$^{-1}$), 
with nothing higher. 

The density contrast between the \ion{H}{2} region electron density $N_e$ 
and the PDR density $n$ is much less than that required 
for pressure equilibrium between a 7000~K \ion{H}{2} region and 
a PDR with $T < 100$~K (e.g., Kaufman et al. 2006, figure~18). 
We investigated this further by computing models 
of a combined \ion{H}{2} region/PDR with Cloudy, where the innermost-zone hydrogen-nucleus density 
is $N_p \sim 316$~cm$^{-3}$ for the \ion{H}{2} region 
and the additional zone densities are computed as required for pressure equilibrium. 
For the default gas-pressure-equilibrium PDR, 
the PDR depth is very narrow, $< 2\arcsec$, in units of arcsec 
scaled to an assumed 8~kpc distance for Sgr~B1 (Figure~8). 
Moreover, the computed PDR density $n$ is over an order of magnitude higher than observed. 
However, when we include additional pressure in the model, 
either turbulent pressure or pressure due to a magnetic field, 
the computed depth of the PDR has values more like the $\sim 1\arcmin$ that is observed, 
and the computed PDR density is also more like the results from the PDR Toolbox, 
$n \sim 1600$~cm$^{-3}$ (Table~2). 
We conclude, in agreement with other studies, 
that significant non-thermal pressure is found in the gas clouds of the GC;  
the result that combined PDR/\ion{H}{2} region models 
are invalid without its inclusion. 

The four massive stars that have been identified in this region 
by Mauerhan et al. (2010) are all either WR or O supergiants. 
We find from the literature that such stars have luminosities of order $\log\ L/L_{\odot} \sim 5.8$ to 6.
The FUV flux from such a single star would produce $G_0 \sim 1200$ 
at a distance $\sim 1\arcmin$, 
in the same range as our estimated $G_0$ from the PDRT.
The sum of the EUV fluxes from these evolved massive stars is approximately half that needed to power the Sgr~B1 \ion{H}{2} region. 
We conclude that Sgr~B1 could be excited by a handful of 
very massive, high luminosity stars widely scattered within its gas clouds. 
We note that such high luminosity but not high temperature stars 
are several million years old and are not indicative of 
currently-forming massive stars, 
unlike the current star formation seen in the nearby region Sgr~B2.
It is not surprising that there are no current clusters of massive star formation 
given the relatively low densities $n$ in the PDR gas. 

An exciting update is that 
in very recent observations, we have identified two additional evolved massive star candidates in Sgr~B1 (these will be presented and discussed in detail in A. Cotera et al., in preparation).   
The ionizing radiation flux from these stars, as described above, makes a significant additional contribution to the energy balance in the gas and dust.   

From the low densities of the PDR (as well as the \ion{H}{2} region) 
and the lack of significant connection 
to the cold dust and molecular gas of the GC, 
we suggest that the ionizing stars of Sgr~B1 represent a currently dispersing OB association with the most massive stars presenting as WR stars. 
This is an interesting contrast to the Arches and Quintuplet Clusters, 
which are dense with stars of similar ages 
but are slowly evaporating in the tidal field of the GC (Habibi et al. 2014).  

We conclude that Sgr~B1 is another example of where high-spatial-resolution 
observations of the GC can inform us of possible conditions 
(e.g., OB associations versus compact star clusters) 
that may be seen in more distant galaxies.  

\vspace{5mm}

\acknowledgments
Based on observations made with the NASA/DLR Stratospheric Observatory for Infrared Astronomy (SOFIA). SOFIA is jointly operated by the Universities Space Research Association, Inc. (USRA), under NASA contract NNA17BF53C, and the Deutsches SOFIA Institut (DSI) under DLR contract 50 OK 0901 to the University of Stuttgart.
Financial support for this work was provided by NASA through awards 04-0113 and 05-0082 issued by USRA. 
We thank Christian Fischer, Randolf Klein, and Bill Vacca for assistance with the observations. 
We thank the referee for the detailed comments that greatly improved the content and the presentation.

\vspace{5mm}
\facility{SOFIA(FIFI-LS)}

\software{Cloudy \citep{cloudy17}, {PDR Toolbox \citep{kwh06},\citep{pound-pdrt}}
}

\end{document}

%% file: tab1.tex
\setcounter{table}{0}
\begin{table}
\centering
\caption{Line Parameters
}
\begin{tabular}{@{}lccc@{}}
\hline
\hline
Line & Wavelength & $IP$\tablenotemark{a} & $E_{upper}$   \\
     & (\micron)  & (eV) & (cm$^{-1}$)                  \\
\hline
H$_2$ S(0) & 28.219  & 0 & 354.37    \\ 
H$_2$ S(1) & 17.035  & 0 & 705.69     \\ 
\mbox{[C\,{\sc ii}]}  $^2$P$_{3/2}-^2$P$_{1/2}$   & 157.741 & 11.26 & 63.40        \\ %
\mbox{[O\,{\sc i}]} $^3$P$_1-^3$P$_2$       & 63.184 & 0 & 158.27        \\ %
\mbox{[O\,{\sc i}]} $^3$P$_0-^3$P$_1$       & 145.525 & 0 & 226.98        \\ %
\mbox{[O\,{\sc iii}]} $^3$P$_1-^3$P$_0$     & 88.356 & 35.12 & 113.18     \\ %
\mbox{[O\,{\sc iii}]} $^3$P$_2-^3$P$_1$     & 51.815 & 35.12 & 306.17     \\ 
\mbox{[Si\,{\sc ii}]} $^2$P$_{3/2}-^2$P$_{1/2}$  & 34.815 & 8.15 & 287.23     \\ %
\mbox{[S\,{\sc iii}]} $^3$P$_1-^3$P$_0$         & 33.481 & 23.34 & 298.68     \\ %
\hline
\end{tabular}
\tablenotetext{a}{Ionization Potential ($IP$) is the energy required to produce the given state of the molecule or ion from the ground state.}
\end{table}



%% file: tab2.tex
\setcounter{table}{1}
\begin{table*}
\centering
\begin{minipage}{164mm}
\caption{Cloudy Models for Sgr B1
}
\begin{tabular}{@{}lccc@{}}
\hline
\hline
Model Parameters\tablenotemark{a}   &     Constant Gas Pressure & Const. Gas+Turb. Pressure &  Const. Gas+Mag.Field Pressure \\
\hline
$V_{turb}$ (km s$^{-1}$)\tablenotemark{b} &  13.54 (no turb. pressure)   &   7.20 (includes turb. pressure)  &   13.54 (no turb. pressure) \\
Magnetic Field $B$ &  0                  &     0               &       $63~\mu$G \\
\hline
Results &&& \\
Separation 17-146\tablenotemark{c} &  $1.2\arcsec$ &  $53.6\arcsec$ & $60.3\arcsec$ \\
{[\ion{S}{3}] 33/[\ion{Si}{2}]} 34 &   2.19    &     3.07     &               2.77 \\
{[\ion{O}{1}] 146/[\ion{C}{2}]} 158  & 0.0889    &     0.01365         &       0.01324 \\
$n$ at 17peak &  44000 cm$^{-3}$         &    1470 cm$^{-3}$        &       1520 cm$^{-3}$ \\
$T$ at 17peak &  110 K              &    94 K             &       81 K \\
$G_0$ at face  &  1900               &    1890             &       2060 \\
\hline
\end{tabular}\
\tablecomments{
\tablenotetext{a}{
Models were computed with Cloudy 17.01 (Ferland et al. 2017). 
Default Cloudy parameters (for details, see Hazy1.pdf, https://www.nublado.org) were used with the following exceptions: 
Model abundances with respect to hydrogen were taken from Simpson (2018): 
(C, N, O, Ne, Si, S, Ar, Fe = 5.13e-4, 1.16e-4, 6.84e-4, 1.74e-4, 2.40e-5, 1.90e-5, 6.2e-6, 2.6e-6) and
`\ion{H}{2} region abundances' of other elements
with twice the default abundances of `ISM grains' and PAHs.
The input stellar spectral energy distribution was a supergiant model atmosphere computed with WM-BASIC with $T_{\rm eff}=32,300$~K, log~$g = 3.23$, from the Binary Population and Spectral Synthesis (BPASS) project (Eldridge et al. 2017). 
The \ion{H}{2} region models were all plane-parallel with ionization parameter log($U) = -2.0$ and filling factor = 0.10, which were determined from a model grid such that the {[\ion{S}{3}] 33/[\ion{Si}{2}]} 34~\micron\ line ratio was in the range of the observed ratios (Simpson et al. 2018).
The input density for combined \ion{H}{2} region/PDR models is the hydrogen nucleus density in the \ion{H}{2} region: 316.2~cm$^{-3}$ (log hden=2.5).
Molecular hydrogen was treated with the `large H2 model' of Shaw et al. (2005).
}
\tablenotetext{b}{Line intensities computed with Cloudy are reduced if the line is optically thick. The reduction can be substantial for lines like the [\ion{C}{2}] 158~\micron\ line, which is very optically thick if there is only thermal line broadening. Additional turbulent broadening can be added to the line widths; it is optional to include it in the pressure. 
For our models, we used either turbulent line broadening $V_{turb} = 13.54$ km~s$^{-1}$ (this gives the hydrogen radio recombination line, RRL, FWHM=32 km~s$^{-1}$, the value found for the brightest RRL by Mehringer et al. 1992) with no additional pressure or 7.20 km~s$^{-1}$ when needed for the additional turbulent pressure model.
 }
\tablenotetext{c}{Separation of the peak emissivities in Figure~8 H$_2$ S(1) 17~\micron\ line (`17peak') and the [\ion{O}{1}] 146~\micron\ line.}
}
\end{minipage}
\end{table*}